\documentclass{article}
\usepackage{epsfig}
\usepackage{color}
\usepackage{amsmath}
\usepackage{amssymb}

\makeatletter
\@addtoreset{equation}{section}

\makeatother

\begin{document}

\begin{flushright}
April 2013

SNUTP13-001
\end{flushright}

\begin{center}

\vspace{3cm}

{\LARGE 
\begin{center}
A Systematic Study on Matrix Models 

for 

Chern-Simons-matter Theories
\end{center}
}

\vspace{2cm}

Takao Suyama \footnote{e-mail address : suyama@phya.snu.ac.kr}

\vspace{1cm}

{\it 
Research Institute of Basic Sciences

and 

Center for Theoretical Physics, 

\vspace{2mm}

Seoul National University, 

Seoul 151-747 Korea}

\vspace{2cm}

{\bf Abstract} 

\end{center}

We investigate the planar solution of matrix models derived from various Chern-Simons-matter theories compatible with the planar limit. 
The saddle-point equations for most of such theories can be solved in a systematic way. 
A relation to Fuchsian systems play an important role in obtaining the planar resolvents. 
For those theories, the eigenvalue distribution is found to be confined in a bounded region even when the 't~Hooft couplings become large. 
As a result, the vevs of Wilson loops are bounded in the large 't~Hooft coupling limit. 
This implies that many of Chern-Simons-matter theories have quite different properties from ABJM theory. 
If the gauge group is of the form ${\rm U}(N_1)_{k_1}\times{\rm U}(N_2)_{k_2}$, then the resolvents can be obtained in a more explicit form than 
in the general cases. 

\newpage

\vspace{1cm}

\section{Introduction}

\vspace{5mm}

Supersymmetric Chern-Simons-matter theories have been intensively studied for several years. 
A recent motivation for studying these theories is to understand the dynamics of M2-branes. 
The worldvolume theory on M2-branes in the flat space-time was constructed in 
\cite{Bagger:2006sk}\cite{Bagger:2007jr}\cite{Bagger:2007vi}\cite{Gustavsson:2008dy}\cite{Gustavsson:2007vu} 
whose applicability is, however, limited to the system with two M2-branes. 
This limitation on the number of M2-branes was lifted in \cite{Aharony:2008ug} in which the background space-time is replaced with an orbifold. 
The worldvolume theory, known as ABJM theory, allows us to take the large $N$ limit, where $N$ is the number of M2-branes. 
In this limit, one finds a relation to M-theory on AdS$_4\times S^7/\mathbb{Z}_k$ \cite{Aharony:2008ug} 
in a quite similar way to the well-studied AdS$_5$/CFT$_4$ correspondence \cite{Maldacena:1997re}. 

Later developments of the study on ABJM theory and other supersymmetric Chern-Simons-matter theories were enabled by 
the derivation of the localization formula for the partition functions and vevs of some BPS operators 
of these theories \cite{Kapustin:2009kz}. 
For ABJM theory, the localization formula 
\begin{equation}
Z_{\rm ABJM}\ \propto\ \int d^N\mu d^N\nu\frac{\prod_{i<j}\sinh^2\frac{\mu_i-\mu_j}2\sinh^2\frac{\nu_i-\nu_j}2}{\prod_{ij}\cosh^2\frac{\mu_i-\nu_j}2}
 \exp\left[ \frac{ik}{4\pi}\sum_{i=1}^N( \mu_i^2-\nu_i^2 ) \right]
   \label{ABJM}
\end{equation}
has been studied in various ways. 
One of the most apparent way to tackle this formula would be 
to regard this integral as a partition function of a matrix model, and to apply the techniques developed 
in the context of matrix models. 
Indeed, such researches appeared soon after the derivation of (\ref{ABJM}) \cite{Suyama:2009pd}\cite{Marino:2009jd}. 
In \cite{Marino:2009jd}, a relation to a topological string theory was also found which turned out to provide a very effective method to derive various results 
\cite{Drukker:2010nc}. 
The power of the relation to the topological string theory was also shown in \cite{Fuji:2011km} which succeeded in deriving 
a surprisingly simple formula for the free energy of 
ABJM theory which includes all $1/N$ corrections. 
The same formula was also derived in \cite{Marino:2011eh} in a different way. 
In \cite{Marino:2011eh}, the partition function (\ref{ABJM}) was related to the thermodynamical partition function of a non-interacting Fermi gas. 
Such a relation to a Fermi gas is familiar in the context of matrix models, but for the case of ABJM theory, the one-particle Hamiltonian is quite non-trivial. 
The relation to a Fermi gas can be extended \cite{Marino:2011eh} to various Chern-Simons-matter theories of quiver types originally studied in 
\cite{Herzog:2010hf}. 
In fact, the localized partition function of a generic Chern-Simons-matter theory can be written as the partition function of a Fermi gas 
\cite{Marino:2012az}, but it is interacting in general, and the analytic power of the reformulation seems to be limited. 
Another type of analysis was developed in \cite{Herzog:2010hf} which restricts oneself to the leading behavior in a particular large $N$ limit. 
(See also \cite{Suyama:2009pd} for a related analysis.) 
It was shown \cite{Herzog:2010hf}\cite{Martelli:2011qj}\cite{Cheon:2011vi} 
that this technique reproduces the volume of the internal manifold in the dual M-theory background. 

\vspace{5mm}

In \cite{Suyama:2012uu}, we investigated the large $N$ solution of a particular family of theories, called Chern-Simons-adjoint theories. 
This family consists of ${\cal N}=3$ ${\rm U}(N)_k$ Chern-Simons theories coupled to arbitrary numbers of adjoint hypermultiplets. 
For most of these theories, the Fermi gas approach \cite{Marino:2011eh} results in interacting theories. 
It was shown in \cite{Suyama:2012uu} that if we restrict ourselves to the planar limit, the solution can be obtained explicitly in terms of theta functions. 
The planar resolvent is given as an integral of a combination of theta functions. 
Although the resulting expression does not look so simple, 
it was shown \cite{Suyama:2012uu} that some properties of the eigenvalue distribution and the behavior of the vev $\langle W \rangle$ 
of a BPS Wilson loop as a function 
of the 't Hooft coupling $t$ can be investigated in some detail. 
For example, $\langle W \rangle$ turned out to be finite even when $t$ diverges. 
This is a quite different behavior from the one which would be expected if there would be a classical AdS dual and $\langle W \rangle$ 
would be given in terms of 
a minimal surface in the AdS space \cite{Rey:1998ik}\cite{Maldacena:1998im}\cite{Rey:1998bq}. 
It would be natural to wonder whether this behavior is special to Chern-Simons-adjoint theories. 
To clarify the situation, it is desired to extend the analysis in \cite{Suyama:2012uu} to more general theories. 

In this paper, we discuss all ${\cal N}=3$ Chern-Simons-matter theories which are compatible with the planar limit. 
Since the constraint due to ${\cal N}=3$ supersymmetry is not so severe, there are huge number of such theories. 
The ability to take the planar limit requires that the representations of the matter multiplets should be small enough. 
It turns out that the analysis of a generic theory is reduced to the one of a related theory including only fundamental and bi-fundamental multiplets, 
under an assumption on eigenvalue distributions based on the numerical observations in \cite{Herzog:2010hf}. 
The planar resolvent of the simpler theory can be written in terms of the solution of a system of differential equations unless the parameters of the theory 
satisfy an algebraic equation. 
Although the solution cannot be written in terms of simple functions, the properties of the solution can be investigated in enough detail. 
In order to relate the problem of solving saddle-point equations to a system of differential equations, the knowledge of the Riemann-Hilbert problem 
in the most basic setup is used. 
We mainly focus on the behavior of $\langle W \rangle$ for large $t$. 
If a Chern-Simons-matter theory has a simple gravity dual in some limit, then $\langle W \rangle$ should exhibit an exponential behavior 
as $t$ becomes large. 
It would be more interesting to investigate, in addition, the behavior of the free energy. 
This will be hopefully discussed elsewhere. 

It turns out that any Chern-Simons-matter theory, which is solvable by the technique developed in this paper, 
exhibits the behavior similar to Chern-Simons-adjoint theories \cite{Suyama:2012uu}. 
This suggests that Chern-Simons-matter theories with simple AdS gravity duals would be quite rare even though there are huge number of 
superconformal Chern-Simons-matter theories. 
A condition is found which shows when a Chern-Simons-matter theory might have more interesting properties. 
Such ``exceptional'' theories include ABJM theory and all other theories with known gravity duals. 
It would be interesting to compare this situation with the one in AdS$_5$/CFT$_4$ correspondence. 
In four-dimensions, superconformal field theories are already rare, but many of them have gravity duals. 

\vspace{5mm}

This paper is organized as follows. 
In section \ref{saddle}, we define a family of theories which is investigated in this paper. 
Starting with the localization formula for the partition function, we derive the saddle-point equations. 
Assuming some symmetry for the eigenvalue distributions, the problem of solving the saddle-point equations for a generic theory is reduced to the one for 
a related theory including only fundamental and bi-fundamental multiplets. 
Unless the theory is ``exceptional'', the saddle-point equations can be further simplified. 
In section \ref{2node}, we focus on theories for which the gauge group is of the form ${\rm U}(N_1)_{k_1}\times {\rm U}(N_2)_{k_2}$. 
The saddle-point equations for these theories can be solved explicitly in terms of theta functions, as in \cite{Suyama:2012uu}. 
The condition for the existence of the solution is investigated, and this turns out to be the condition for a divergenct behavior of the 't~Hooft couplings. 
In section \ref{reform}, we solve the saddle-point equations for more general theories. 
It turns out that the technique used in section \ref{2node} cannot be straightforwardly extended to the general cases. 
Instead, the saddle-point equations can be solved by relating them to the solution of a Fuchsian system, a set of first-order differential equations. 
The relation to the Fuchsian system is given by the known results on the Riemann-Hilbert problem. 
Using the structure of the solution of the Fuchsian system, it is possible to extract some properties of the eigenvalue distributions. 
A relation between the divergent behavior of the 't~Hooft couplings and the isomonodromic deformation of the Fuchsian system is pointed out. 
Section \ref{discuss} is devoted to discussion. 
Appendices contain some technical details.

\vspace{1cm}

\section{Planar limit of ${\cal N}=3$ Chern-Simons-matter theories} \label{saddle}

\vspace{5mm}

A family ${\cal C}$ of theories discussed in this paper consists of ${\cal N}=3$ Chern-Simons-matter theories which are compatible with the planar limit. 
In this section, we describe the family ${\cal C}$, and for each theory in ${\cal C}$, 
derive the equations which determine the saddle-point of the localized partition function. 
Some physical quantities can be obtained from the solution of the saddle-point equations. 
It will turn out that the study of a generic theory in ${\cal C}$ in the planar limit is essentially reduced to the study of a theory in a 
sub-family ${\cal C}_0$ 
in which the theories consist of only fundamental and bi-fundamental multiplets. 

\vspace{5mm}

\subsection{Saddle-point equations for theories in ${\cal C}$}

\vspace{5mm}

A theory in ${\cal C}$ is an ${\cal N}=3$ Chern-Simons theory coupled to various matter multiplets. 
The gauge group $G$ of the theory is a direct product of several unitary groups, 
\begin{equation}
G\ =\ \prod_{a=1}^{n_g}{\rm U}(N_a)_{k_a}, 
\end{equation}
where $k_a$ is the Chern-Simons level for the gauge group factor ${\rm U}(N_a)$. 
The gauge fields with their superpartners form ${\cal N}=4$ vector multiplets, and 
the matter fields form ${\cal N}=4$ hypermultiplets. 
The Lagrangians of Chern-Simons-matter theories with ${\cal N}\ge4$ supersymmetry were constructed in 
\cite{Gaiotto:2008sd}\cite{Hosomichi:2008jd}\cite{Aharony:2008ug}\cite{Hosomichi:2008jb}. 
For a generic choice of $G$ and matter multiplets, a Chern-Simons-matter theory possesses at most ${\cal N}=3$ supersymmetry 
since the Chern-Simons terms break the ${\cal N}=4$ supersymmetry partially. 
For an ${\cal N}=3$ theory, 
the superpotential is uniquely determined by the symmetry. 
Therefore, 
the Lagrangian of an ${\cal N}=3$ theory is uniquely specified by 
the choice of the gauge group $G$ including the Chern-Simons levels $\{k_a\}$ and the representations of $G$ to which the matter multiplets belong. 
One can find more details on supersymmetric Chern-Simons-matter theories in \cite{Gaiotto:2007qi}. 

Note that, for some choices of the data mentioned above,  it is argued that the corresponding ${\cal N}=3$ Chern-Simons-matter theories do not exist 
as consistent quantum field theories. 
An example is ABJ theory \cite{Aharony:2008gk}, regarded as an ${\cal N}=3$ theory, with the gauge group ${\rm U}(N_1)_k\times {\rm U}(N_2)_{-k}$ where 
the parameters satisfy $|N_1-N_2|>k$. 
We assume that such theories are excluded from ${\cal C}$. 
It should be noted that the criterion for selecting the data which 
provide us with a consistent theory does not seem to 
be fully investigated. 
This issue will be discussed elsewhere. 

As mentioned above, each theory in ${\cal C}$ is further restricted such that it is compatible with the planar limit. 
The possible representations of $G$ allowed for theories in ${\cal C}$ are the followings, 
\begin{itemize}
\item a fundamental multiplet ({\bf f}) in $N_a\oplus \overline{N_a}$ of ${\rm U}(N_a)$, 
\item an adjoint multiplet ({\bf ad}) in ${\bf adj}\oplus{\bf adj}$ of ${\rm U}(N_a)$, 
\item a symmetric multiplet ({\bf s}) in ${\bf sym}\oplus\overline{\bf sym}$ of ${\rm U}(N_a)$, 
\item an anti-symmetric multiplet ({\bf as}) in ${\bf asym}\oplus\overline{\bf asym}$ of ${\rm U}(N_a)$, 
\item a double-fundamental multiplet ({\bf ff}) in $(N_a\otimes N_b)\oplus(\overline{N_a}\oplus \overline{N_b})$ of ${\rm U}(N_a)\times{\rm U}(N_b)$, 
\item a bi-fundamental multiplet ({\bf bf}) in $(N_a\otimes \overline{N_b})\oplus(\overline{N_a}\oplus N_b)$ of ${\rm U}(N_a)\times{\rm U}(N_b)$. 
\end{itemize}
Note that an ${\cal N}=4$ hypermultiplet consists of two ${\cal N}=2$ chiral multiplets in the representations $R$ and $\overline{R}$. 

\begin{figure}[tbp]
\begin{center}
\includegraphics{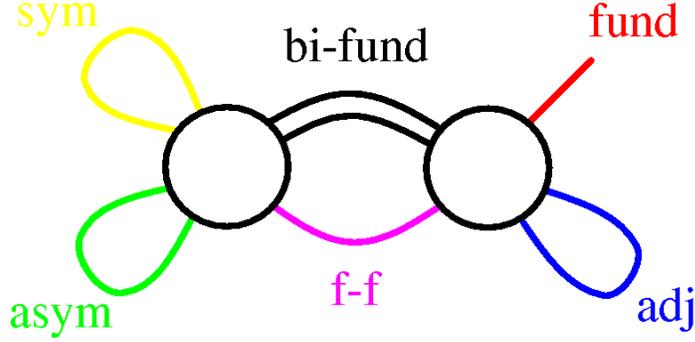}
\end{center}
\caption{
An example of a diagram with two nodes. Various edges correspond to matters in various representations. 
}
\label{diagram}
\end{figure}

Each theory in ${\cal C}$ is associated with a diagram $\Gamma$ which consists of $n_g$ nodes corresponding to ${\rm U}(N_a)$ factors 
and a number of colored edges each of which corresponds to one matter multiplet. 
An edge with a free end corresponds to {\bf f}. 
An edge connected to the same node corresponds to either {\bf ad}, {\bf s} or {\bf as} which are distinguished by a suitable choice of colors. 
An edge connected to two different nodes corresponds to either {\bf ff} or {\bf bf}. 
An example of such a diagram is shown in Figure \ref{diagram}. 
Since the hypermultiplets are non-chiral, there is no natural rule to assign directions to the edges. 
In other words, $\Gamma$ is not a quiver diagram. 
In the following, a theory in ${\cal C}$ associated to a diagram $\Gamma$ will be often called a theory $\Gamma$. 

\vspace{5mm}

For a given theory $\Gamma$, one may calculate the sphere partition function $Z(\Gamma)$, that is, the partition 
function of the theory $\Gamma$ defined on the Euclidean three-sphere $S^3$. 
Using the supersymmetric localization \cite{Kapustin:2009kz}, $Z(\Gamma)$ can be written in terms of a finite-dimensional Riemann integral. 
The integral is taken over the eigenvalues $\{u^a_{i_a}\}$ ($i_a=1,\cdots,N_a$) 
of $n_g$ matrices $\{\sigma^a\}$ where each $\sigma^a$ is an $N_a\times N_a$ Hermitian matrix. 
The explicit formula for $Z(\Gamma)$, with the overall numerical constant omitted, is as follows, 
\begin{eqnarray}
Z(\Gamma) &=& \int \prod_{a=1}^{n_g}\prod_{i_a=1}^{N_a}du^{a}_{i_a}\,\exp\left( -\sum_{a=1}^{n_g}S^a_{\rm tree}[u]-\sum_{a=1}^{n_g}S^a_{\bf v}[u]-S_m[u] \right), 
   \label{localizedZ}
\end{eqnarray}
where $S^a_{\rm tree}[u]$ comes from the classical action, $S^a_{\bf v}[u]$ is the contributions from the $a$-th vector multiplet and 
\begin{eqnarray}
S_m[u] 
&=& \sum_{a=1}^{n_g}\Bigl( n^a_{\bf f}S^a_{\bf f}[u]+n^a_{\bf ad}S^a_{\bf ad}[u]
      +n^a_{\bf s}S^a_{\bf s}[u]+n^a_{\bf a}S^a_{\bf as}[u] \Bigr) \nonumber \\
& & +\sum_{a,b=1}^{n_g}\Bigl( n^{ab}_{\bf ff}S^{ab}_{\bf ff}[u]+n^{ab}_{\bf bf}S^{ab}_{\bf bf}[u] \Bigr)
\end{eqnarray}
is a sum of contributions from the matter multiplets. 
Here $n^a_{R}$ and $n^{ab}_R$ are the numbers of the matter multiplets in the representation $R$ coupled to ${\rm U}(N_a)$ and 
${\rm U}(N_a)\times {\rm U}(N_b)$, respectively. 
The details of those terms are as follows, 
\begin{equation}
\begin{array}{lcl}
\displaystyle{S^a_{\rm tree}[u] \ =\  \sum_{i_a}\frac{k_a}{4\pi i}(u^a_{i_a})^2}, & \hspace{5mm} & 
 \displaystyle{S^a_{\bf v}[u] \ =\  -\sum_{i_a<j_a}\log\left[ \sinh^2\frac{u^a_{i_a}-u^a_{j_a}}2 \right]}, \nonumber \\ [5mm] 
\displaystyle{S^a_{\bf f}[u] \ =\  \sum_{i_a}\log\left[ \cosh\frac{u^a_{i_a}}2 \right]}, & \hspace{5mm} & 
 \displaystyle{S^a_{\bf ad}[u] \ =\  \sum_{i_a\ne j_a}\log\left[ \cosh\frac{u^a_{i_a}-u^a_{j_a}}2 \right]}, \nonumber \\ [5mm] 
\displaystyle{S^a_{\bf s}[u] \ =\  \sum_{i_a\le j_a}\log\left[ \cosh\frac{u^a_{i_a}+u^a_{j_a}}2 \right]}, & \hspace{5mm} & 
 \displaystyle{S^a_{\bf as}[u] \ =\  \sum_{i_a< j_a}\log\left[ \cosh\frac{u^a_{i_a}+u^a_{j_a}}2 \right]}, \nonumber \\ [5mm] 
\displaystyle{S^{ab}_{\bf ff}[u] \ =\  \sum_{i_a,j_b}\log\left[ \cosh\frac{u^a_{i_a}+u^b_{j_b}}2 \right]}, & \hspace{5mm} & 
 \displaystyle{S^{ab}_{\bf bf}[u] \ =\  \sum_{i_a,j_b}\log\left[ \cosh\frac{u^a_{i_a}-u^b_{j_b}}2 \right]}. 
\end{array}
\end{equation}

In the following, we replace $S^a_{\rm tree}[u]$ with 
\begin{equation}
\tilde{S}^a_{\rm tree}[u] \ =\  \sum_{i_a}\frac{k_a}{4\pi}(u^a_{i_a})^2, 
\end{equation}
and assume that all $k_a$ are real and positive. 
This is sufficient to make the integral (\ref{localizedZ}) well-defined for any choice of matter multiplets. 
The partition function of the original theory will be recovered via a suitable analytic continuation of $\{k_a\}$ from real positive values to purely imaginary 
values. 

\vspace{5mm}

Consider a limit in which $\{N_a\}$ and $\{k_a\}$ are all proportional to a common number, say $k$, and $k$ becomes infinitely large. 
The proportionality constants are denoted as 
\begin{equation}
t^a\ :=\ \frac{2\pi N_a}{k}, \hspace{5mm} \kappa^a\ :=\ \frac{k_a}{k}, 
\end{equation}
where $\{t^a\}$ are the 't~Hooft couplings of the Chern-Simons-matter theory. 
If there are a number of {\bf f}s, $\{n^a_{\bf f}\}$ are also assumed to be proportional to $k$. 
As is familiar in the context of matrix models, the integral $Z(\Gamma)$ is dominated by a particular configuration of the eigenvalues $\{u^a_{i_a}\}$ in 
this limit. 
The configuration is determined by the saddle-point equations 
\begin{equation}
\frac{k_a}{2\pi}u^a_{i_a}\ =\ \sum_{j_a\ne i_a}\coth\frac{u^a_{i_a}-u^a_{j_a}}2-\frac{\partial S_m[u]}{\partial u^a_{i_a}}. 
   \label{SP_u}
\end{equation}

Let $\bar{u}$ denote a solution of these equations. 
The free energy $F(\Gamma)$ of the theory $\Gamma$ in the planar limit is given as 
\begin{equation}
F(\Gamma)\ =\ \sum_{a=1}^{n_g}S^a_{\rm tree}[\bar{u}]+\sum_{a=1}^{n_g}S^a_{\bf v}[\bar{u}]+S_m[\bar{u}]. 
\end{equation}
The leading term of $F(\Gamma)$ in the planar limit is proportional to $k^2$, and the coefficient is a function of the 't~Hooft 
couplings $\{t^a\}$. 
Some other observables consistent with the localization can be also determined in terms of $\bar{u}$. 
For example, the expectation value of a supersymmetric Wilson loop for a ${\rm U}(N_a)$ factor is given as 
\begin{equation}
\langle W^a \rangle\ =\ \frac1{N_a}\sum_{i_a=1}^{N_a}\exp\left(  \bar{u}^a_{i_a} \right). 
\end{equation}

The theory $\Gamma$ may have interesting properties, for example, the existence of a classical gravity dual. 
Typically, such a dual description appears when (some of) $\{t^a\}$ become large. 
Since the solution $\bar{u}$ is determined once $\{t^a\}$ are given (it may not be unique, see section \ref{2node}), 
one can regard $\langle W^a \rangle$ as a function of $\{t^a\}$. 
If $\langle W^a \rangle$ would be given in terms of a minimal surface in the dual geometry (possibly an AdS space), then $\langle W^a \rangle$ would grow 
exponentially as $\{t^a\}$ grow. 
Therefore, the behavior of $\langle W^a \rangle$ when $\{t^a\}$ diverge indicates whether the theory $\Gamma$ is interesting in the above sense.

\vspace{5mm}

\subsection{Planar relations in ${\cal C}$}  \label{equivalence}

\vspace{5mm}

The family ${\cal C}$ defined so far consists of various kinds of Chern-Simons-matter theories. 
The variety looks so rich that one might think one needs to investigate those theories one by one. 
Fortunately, it turns out that they can be investigated in a rather systematic way under an assumption on the eigenvalue distributions. 

In \cite{Herzog:2010hf}, the eigenvalue distributions were numerically determined for some matrix models of Chern-Simons-matter theories. 
The plots in \cite{Herzog:2010hf} shows that the distributions are invariant under the reflection 
\begin{equation}
u^a_{i_a}\ \to\ -u^a_{i_a}. 
   \label{reflection}
\end{equation}
In fact, this reflection, performed simultaneously on all eigenvalues, 
is a symmetry of the saddle-point equations (\ref{SP_u}), and therefore, it would be natural to assume 
that the solution should be also symmetric. 
In the following, we assume that 
\begin{equation}
\{\,u^a_1,\cdots,u^a_{N_a}\,\}\ =\ \{\,-u^a_1,\cdots,-u^a_{N_a}\,\}
\end{equation}
holds as sets for any $a=1,\cdots,n_g$. 
Under this assumption, it turns out that different Chern-Simons-matter theories share the same saddle-point equations, and 
therefore the same eigenvalue distributions. 

For example, the contribution to (\ref{SP_u}) from {\bf ff}, 
\begin{equation}
-\sum_{j_b}\frac{n^{ab}_{\bf ff}}2\tanh\frac{u^a_{i_a}+u^b_{j_b}}2, 
\end{equation}
coincides with the contribution from {\bf bf}, 
\begin{equation}
-\sum_{j_b} \frac{n^{ab}_{\bf bf}}2\tanh\frac{u^a_{i_a}-u^b_{i_b}}2, 
   \label{bf}
\end{equation}
with $n^{ab}_{\bf bf}=n^{ab}_{\bf ff}$. 
Therefore, as long as the planar solution is concerned, one may replace {\bf ff} with the same number of {\bf bf} 
and obtain the same eigenvalue distributions. 
In this sense, it is not necessary to discuss any theory $\Gamma$ including {\bf ff}. 

By the same token, the matters {\bf s} and {\bf as}, whose contributions are 
\begin{equation}
\left[ -\frac{n^a_{\bf s}}2\sum_{j_a}\tanh\frac{u^a_{i_a}+u^a_{j_a}}2-\frac{n^a_{\bf s}}2\tanh u^a_{i_a} \right]
 +\left[ -\frac{n^a_{\bf as}}2\sum_{j_a}\tanh\frac{u^a_{i_a}+u^a_{j_a}}2+\frac{n^a_{\bf as}}2\tanh u^a_{i_a} \right],
\end{equation}
can be replaced with {\bf ad}, which contributes as 
\begin{equation}
-n^a_{\bf ad}\sum_{j_a}\tanh\frac{u^a_{i_a}-u^a_{j_a}}2,
   \label{ad}
\end{equation}
with $n^a_{\bf ad}=\frac12n^a_{\bf s}+\frac12n^a_{\bf as}$, up to ``local terms'' 
\begin{equation}
-\frac{n^a_{\bf s}-n^a_{\bf as}}2\tanh u^a_{i_a}. 
\end{equation}
In fact, it can be shown that these ``local terms'' can be ignored in the planar limit. 
If $n^a_{\bf s}$ and $n^a_{\bf as}$ are of order $k^0$, then this term is smaller than the other terms in (\ref{SP_u}) by $k^{-1}$. 
On the other hand, if either $n^a_{\bf s}$ or $n^a_{\bf as}$ is of order $k$, then the term (\ref{ad}) becomes larger than the other terms by $k$, and 
the equations (\ref{SP_u}) do not have a non-trivial solution. 

\begin{figure}[tbp]
\begin{center}
\includegraphics{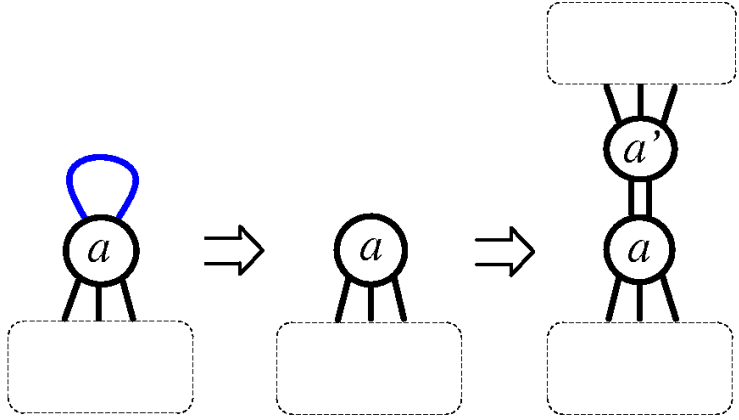}
\end{center}
\caption{
The procedure for replacing {\bf ad} with {\bf bf}. 
}
\label{ad2bf}
\end{figure}

There is a further simplification in which {\bf ad} can be replaced with {\bf bf} by changing the diagram $\Gamma$. 
This possibility can be anticipated from the similarity between (\ref{bf}) and (\ref{ad}). 
Suppose that, in a theory $\Gamma$, there is an $a$-th node with edges for {\bf ad}. 
The replacement goes as follows. 
One eliminates the edges in $\Gamma$ corresponding to {\bf ad} and obtains a diagram $\Gamma'$. 
Then one duplicates $\Gamma'$ 
and connects the $a$-th node in the original $\Gamma'$ with the same node in its mirror $\Gamma'$ by the edges corresponding to {\bf bf}. 
The number of edges for {\bf bf} is the twice the number of {\bf ad} which originally existed. 
See Figure \ref{ad2bf}. 
Denote the resulting diagram as $\Gamma''$. 
The saddle-point equations (\ref{SP_u}) for the theory $\Gamma''$ are no longer equivalent to the original ones. 
However, they are equivalent to the original equations for the theory $\Gamma$ if the mirror pairs of nodes in $\Gamma''$ are assumed to have the 
same eigenvalue distributions. 
In this way, the planar solution of the original theory $\Gamma$ can be obtained from the solution of the enlarged theory $\Gamma''$ by suitably 
constraining the eigenvalue distributions. 
Repeating this procedure, one can find a diagram $\Gamma'''$ including only edges for {\bf f} and {\bf bf} from which one obtains the planar solution of the 
original theory $\Gamma$ including {\bf ad}. 

\vspace{5mm}

Now, the problem of solving saddle-point equations (\ref{SP_u}) for a generic theory in ${\cal C}$ has been reduced to solving 
\begin{equation}
\frac{k_a}{2\pi}u^a_{i_a}+\frac{n^a}2\tanh\frac{u^a_{i_a}}2
 \ =\ \sum_{j_a\ne i_a}\coth\frac{u^a_{i_a}-u^a_{j_a}}2-\sum_b\sum_{j_b}\frac{n^{ab}}2\tanh\frac{u^a_{i_a}-u^b_{j_b}}2, 
   \label{SP_reduced}
\end{equation}
where $n^a,n^{ab}$ are integers. 
Note that $n^a,n^{ab}$ are always non-negative. 
These are the saddle-point equations for a theory $\Gamma$ including only edges for {\bf f} and {\bf bf}, that is, the equations 
for a theory in ${\cal C}_0$. 
Note that the analysis below shows that $n^a,n^{ab}$ can take any values other than integers as long as the equations (\ref{SP_reduced}) are concerned. 
In fact, there exists a solution even when they are complex numbers.

\vspace{5mm}

\subsection{Equations in terms of resolvents}

\vspace{5mm}

To study the equations (\ref{SP_reduced}) further, 
it is convenient to introduce new variables 
\begin{equation}
z^a_{i_a}\ :=\ \exp\left( u^a_{i_a} \right). 
\end{equation}
The reflection symmetry (\ref{reflection}) of $u^a_{i_a}$ is translated into the symmetry under the inversion 
\begin{equation}
z^a_{i_a}\ \to\ (z^a_{i_a})^{-1}. 
   \label{reflection_z}
\end{equation}
In terms of $z^a_{i_a}$, the equations (\ref{SP_reduced}) can be written as 
\begin{equation}
\frac{k_a}{2\pi}\log z^a_{i_a}+\frac{n^a}2\frac{z^a_{i_a}-1}{z^a_{i_a}+1}
 \ =\ \sum_{j_a\ne i_a}\frac{z^a_{i_a}+z^a_{j_a}}{z^a_{i_a}-z^a_{j_a}}-\sum_b\frac{n^{ab}}2\sum_{j_b}\frac{z^a_{i_a}-z^b_{j_b}}{z^a_{i_a}+z^b_{j_b}}. 
   \label{SP_z}
\end{equation}

A standard strategy to deal with this kind of equations is to rewrite them in terms of resolvents defined as 
\begin{equation}
v^a(z)\ :=\ t^a\cdot\frac1{N_a}\sum_{j_a}\frac{z+z^a_{j_a}}{z-z^a_{j_a}}. 
\end{equation}
The 't~Hooft coupling $t^a$ can be easily read off from $v^a(z)$ as 
\begin{equation}
t^a\ =\ -v^a(0)\ =\ v^a(\infty). 
\end{equation}
The vev $\langle W^a \rangle$ can be also obtained from the expansion of $v^a(z)$ for large $z$, 
\begin{equation}
v^a(z) =\ t^a\left[ 1+2\langle W^a \rangle z^{-1}+O(z^{-2}) \right]. 
\end{equation}
Note that the symmetry under (\ref{reflection_z}) implies 
\begin{equation}
v^a(z^{-1})\ =\ -v^a(z). 
   \label{inversion}
\end{equation}

In the planar limit, it is assumed that, for each $a$, the eigenvalues $\{z^a_{i_a}\}$ condense to form a continuous distribution whose support is an 
interval $[p_a,q_a]$, and $v^a(z)$ becomes a holomorphic function on $\mathbb{P}^1\backslash[p_a,q_a]$ having a branch cut on $[p_a,q_a]$. 
Note that the symmetry under the inversion (\ref{reflection_z}) implies 
\begin{equation}
p_aq_a\ =\ 1 
\end{equation}
for each $a$. 
We further assume that $v^a(z)$ is finite at the branch points. 
This is based on the integral expression of $v^a(z)$ 
\begin{equation}
v^a(z)\ = \ \int_{p_a}^{q_a}dx\,\rho^a(x)\frac{z+x}{z-x}, 
\end{equation}
where $\rho^a(x)$ is assumed to be continuous in the planar limit. 
As long as $\rho^a(x)$ decays near $x=p_a$ as a positive power of $x-p_a$, then $v^a(z)$ is finite at $z=p_a$. 

\begin{figure}[tbp]
\begin{center}
\includegraphics{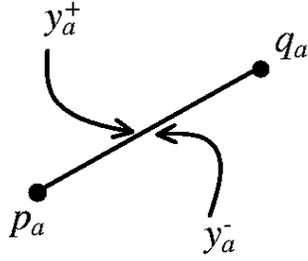}
\end{center}
\caption{
Two limits $y^\pm_a$ to approach $y_a$. 
}
\label{y^+-}
\end{figure}

The right-hand side of (\ref{SP_z}) can be written in terms of $v^a(z)$ as 
\begin{equation}
\frac12\left[ v^a(y_a^+)+v^a(y_a^-)-\sum_bn^{ab}v^b(-y_a) \right]
   \label{non-local}
\end{equation}
where $y_a\in[p_a,q_a]$. 
The quantities $y_a^\pm$ are defined to be $y_a\pm i0$ if the interval $[p_a,q_a]$ lies on the real axis. 
Otherwise, $y_a^\pm$ are defined as limits as depicted in Figure \ref{y^+-}. 
The expression (\ref{non-local}) is, however, not useful to be dealt with since it is ``non-local'', that is, the functions in (\ref{non-local}) 
depend on both $y_a$ and $-y_a$. 
A better expression can be obtained by introducing another set of resolvents 
\begin{equation}
\tilde{v}^a(z)\ :=\ t^a\cdot\frac1{N_a}\sum_{j_a}\frac{z-z^a_{j_a}}{z+z^a_{j_a}}. 
\end{equation}
Note that 
\begin{equation}
\tilde{v}^a(z)=v^a(-z)
   \label{b and w}
\end{equation}
holds. 
Then, obviously (\ref{non-local}) becomes 
\begin{equation}
\frac12\left[ v^a(y_a^+)+v^a(y_a^-)-\sum_bn^{ab}\tilde{v}^b(y_a) \right] 
\end{equation}
which looks ``local''. 
This is also equivalent to another ``local'' expression 
\begin{equation}
\frac12\left[ \tilde{v}^a(-y_a^+)+\tilde{v}^a(-y_a^-)-\sum_bn^{ab}v^b(-y_a) \right]. 
\end{equation}

These expressions suggest that the equations (\ref{SP_z}) can be written in ``local'' forms if one can choose either $v^a(z)$ or $\tilde{v}^a(z)$ for each 
node consistently such that any edge in the diagram $\Gamma$ connects a node for which $v^a(z)$ is chosen and another node for which $\tilde{v}^a(z)$ 
is chosen. 
This choice is possible if and only if the diagram $\Gamma$ is bipartite. 
In this case, one may choose $v^a(z)$ for white nodes and $\tilde{v}^b(z)$ for black nodes. 
Then the equations (\ref{SP_z}) can be written as 
\begin{eqnarray}
2\kappa^a\log y_a+\frac{n^a}k\frac{y_a-1}{y_a+1}
&=& v^a(y_a^+)+v^a(y_a^-)-\sum_bn^{ab}\tilde{v}^b(y_a), 
   \label{SP_inhomogeneous1} \\
2\kappa^b\log (-y_b)+\frac{n^b}k\frac{-y_b-1}{-y_b+1}
&=& \tilde{v}^b(y_b^+)+\tilde{v}^b(y_b^-)-\sum_an^{ba}v^a(y_b),  
   \label{SP_inhomogeneous2}
\end{eqnarray}
where $a$ runs over white nodes and $b$ runs over black nodes, and $y^b\in[-q_b,-p_b]$. 
Since, for each $a$, either $v^a(z)$ or $\tilde{v}^a(z)$ appears in the above equations, one may forget the constraints (\ref{b and w}). 

\begin{figure}[tbp]
\begin{center}
\includegraphics{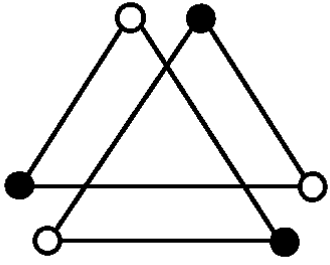}
\end{center}
\caption{
The bipartite double cover of a triangle. 
}
\label{double cover}
\end{figure}

In the case when $\Gamma$ is not bipartite, for example when $\Gamma$ is a triangle, one may obtain a bipartite diagram from $\Gamma$ 
by duplicating the nodes and 
draw edges such that each edge connects a node in the original diagram and another node in the mirror. 
See Figure \ref{double cover}. 
The resulting diagram $\tilde{\Gamma}$, known as the bipartite double cover of $\Gamma$, is always bipartite by construction. 
The saddle-point equations for $\tilde{\Gamma}$ are nothing but (\ref{SP_inhomogeneous1})(\ref{SP_inhomogeneous2}) with the indices $a,b$ both run 
over all nodes of $\Gamma$. 
One may obtain the resolvents of the original theory from the solution of the equations (\ref{SP_inhomogeneous1})(\ref{SP_inhomogeneous2}) 
by imposing the constraint (\ref{b and w}). 

In the following, the diagram $\Gamma$ is always assumed to be bipartite. 
The equations (\ref{SP_inhomogeneous1})(\ref{SP_inhomogeneous2}) can be written simply as 
\begin{equation}
2\kappa^a\log (\epsilon(a)y_a)+\frac{n^a}k\frac{\epsilon(a)y_a-1}{\epsilon(a)y_a+1} \ =\  v^a(y_a^+)+v^a(y_a^-)-\sum_bn^{ab}v^b(y_a), 
   \label{SP_bipartite}
\end{equation}
where the branch cut of $v^a(z)$ is on the interval $[p_a,q_a]$ and 
\begin{equation}
\epsilon(a)\ =\ \left\{ 
\begin{array}{cc}
+1, & (a\mbox{ is a white node,}) \\
-1, & (a\mbox{ is a black node.})
\end{array}
\right. 
\end{equation}

\vspace{5mm}

\subsection{Integral ansatz} \label{homogeneous}

\vspace{5mm}

The equations (\ref{SP_bipartite}) look like the ones encountered in the context of matrix models, except for 
the logarithmic functions in the left-hand side. 
They have a branch cut, which would make the procedure of solving the equations, based on the analyticity, more difficult than the usual cases 
where only polynomial functions appear. 

It was found in \cite{Suyama:2012uu} that the following ansatz 
\begin{equation}
v^a(z)\ =\ \int_Cd\xi\,v^a(z,\xi)
   \label{ansatz}
\end{equation}
for the resolvents is quite useful to deal with the logarithmic functions. 
Here the integration contour $C$ starts from the origin of the complex plane $\mathbb{C}$ and reaches infinity, avoiding the branch cut of 
$v^a(z)$. 

The equations (\ref{SP_bipartite}) can be reduced to a set of homogeneous equations using (\ref{ansatz}) as follows. 
Let us first consider the case $n^a=0$. 
Suppose that a set of functions $\{v^a(z,\xi)\}$ satisfies the equations 
\begin{equation}
f^a(y_a,\xi)\ =\ v^a(y_a^+,\xi)+v^a(y_a^-,\xi)-\sum_bn^{ab}v^b(y_a,\xi)
   \label{SP_integrand}
\end{equation}
where 
\begin{equation}
f^a(z,\xi)\ =\ -\frac{2\kappa^a}{\epsilon(a)z-\xi}-\frac{2\kappa^a}{\xi-1}
\end{equation}
are rational functions of $z$. 
Integrating both sides of the equations (\ref{SP_integrand}), one recovers the original equations (\ref{SP_bipartite}). 

The integrand $v^a(z,\xi)$ is assumed to be holomorphic on $\mathbb{P}^1\backslash[p^a,q^a]$, to have a branch cut on $[p^a,q^a]$ 
for all values of $\xi$ and to be finite at the branch points. 
As long as the integral (\ref{ansatz}) is well-defined, the function $v^a(z)$ obtained from the solution of (\ref{SP_integrand}) 
is also holomorphic on $\mathbb{P}^1\backslash[p^a,q^a]$ having a branch cut on $[p^a,q^a]$, and finite at the branch points. 

To eliminate the rational function $f^a(z,\xi)$, let us define a complex function $\omega^a(z,\xi)$ such that 
\begin{equation}
v^a(z,\xi)\ =\ r^a(z,\xi)+\omega^a(z,\xi) 
   \label{rational part}
\end{equation}
holds for some rational function $r^a(z,\xi)$. 
If the following set of linear equations 
\begin{equation}
f^a(z,\xi)\ = \ \sum_b(2\delta^{ab}-n^{ab})r^a(z,\xi)
   \label{linear_eq}
\end{equation}
has the solution, then the equations (\ref{SP_integrand}) are reduced to the following homogeneous equations 
\begin{equation}
\omega^a(y_a^+,\xi)+\omega^a(y_a^-,\xi)-\sum_bn^{ab}\omega^b(y_a)\ =\ 0. 
   \label{SP_homogeneous}
\end{equation}
The function $\omega^a(z,\xi)$ must have appropriate poles which cancel those of $r^a(z,\xi)$ such that $v^a(z,\xi)$ does not have such poles. 
Note that the introduction of poles is much easier to be dealt with than that of log-cut, which would be necessary at the level of $v^a(z)$. 
This is an advantage of the integral ansatz (\ref{ansatz}). 

In the case $n^a>0$, the reduction may proceed in two steps. 
Since the terms in (\ref{SP_integrand}) proportional to $n^a$ are rational, they can be eliminated by assuming 
\begin{equation}
v^a(z)\ =\ r^a_1(z)+v^a_1(z) 
   \label{fund}
\end{equation}
where $\{r^a_1(z)\}$ are rational functions satisfying the equations similar to (\ref{linear_eq}). 
Then, employing the integral ansatz, the remaining logarithmic terms can be eliminated as above. 

\vspace{5mm}

If the equations (\ref{SP_homogeneous}) could be solved, then the 't~Hooft coupling $t^a$ could be obtained from $\omega^a(z,\xi)$ as 
\begin{equation}
t^a\ =\ -\int_Cd\xi\left[ r^a(0,\xi)+\omega^a(0,\xi) \right]. 
   \label{'t Hooft}
\end{equation}
This equation determines $t^a$ as a function of the positions $\{p_a,q_a\}$ of the branch cuts. 
The analytic continuation of $t^a$ can be done by changing the configuration of the branch cuts. 

Note that the parameters which are given by specifying a diagram $\Gamma$ are the 't~Hooft couplings $\{t^a\}$, while 
the positions $\{p_a,q_a\}$ of the branch cuts would be determined according to the relations (\ref{'t Hooft}). 
Later, it will turn out that the relations (\ref{'t Hooft}) for all $a$ do not always define a one-to-one map between the two parameter spaces, even when 
$p_aq_a=1$ is imposed for any $a$. 
This implies that the saddle-point is not always unique for a given set of $\{t^a\}$. 
In the case there are multiple saddle-points, it is necessary to compare the values of the free energy at those saddle-points 
to correctly extract physical information. 
This issue will not be investigated further in this paper. 

Now, the problem is reduced to solving the above homogeneous equations (\ref{SP_homogeneous}). 
In the next section, we investigate the simplest non-trivial case in which the diagram $\Gamma$ consists of two nodes. 
Theories with more nodes will be discussed in section \ref{reform}. 

\vspace{5mm}

It was found above that if 
\begin{equation}
\mbox{det}(2\delta^{ab}-n^{ab})\ =\ 0
   \label{exception}
\end{equation}
is satisfied for a theory $\Gamma$, then the technique which will be developed in later sections cannot be applied to the theory. 
Indeed, since the left-hand side of (\ref{linear_eq}) is a set of functions, they cannot be always in the image of a degenerate 
constant matrix $2\delta^{ab}-n^{ab}$. 
We call a theory $\Gamma$ in ${\cal C}_0$ satisfying (\ref{exception}) a {\it degenerate} theory. 
Interestingly, many Chern-Simons-matter theories studied so far are degenerate. 
The examples include ABJM theory \cite{Aharony:2008ug}\cite{Aharony:2008gk} 
and its orbifolds \cite{Terashima:2008ba}, circular quivers \cite{Jafferis:2008qz}, necklace quivers \cite{Herzog:2010hf} and more general quivers 
\cite{Gulotta:2011vp}\cite{Gulotta:2012yd}. 
These theories have been already solved by various techniques. 
Especially, some of them are investigated in detail using the Fermi gas approach \cite{Marino:2011eh}. 
See also \cite{Hatsuda:2012dt}\cite{Calvo:2012du}\cite{Hatsuda:2013gj} for recent achievements on instanton effects in this direction. 
It seems that those techniques are not so much useful to generic non-degenerate theories in ${\cal C}_0$, 
and therefore, the technique in this paper would be complementary to those known techniques. 
It is interesting to classify the solutions of (\ref{exception}) and clarify what kinds of theories are included in the class of degenerate theories. 
The results obtained below for non-degenerate theories in ${\cal C}_0$ seem to suggest that the theories which have their simple 
gravity duals would be always degenerate. 

An example of an infinite family of solutions of (\ref{exception}) which do not correspond to the theories mentioned above is, for $n_g=4$,  
\begin{equation}
n^{12}\ =\ n^{21}\ =\ n^{34}\ =\ n^{43}\ =\ 2l, \hspace{5mm} n^{23}\ =\ n^{32}\ =\ 2l^2-2
\end{equation}
and other components of $n^{ab}$ are zero, for any positive integer $l$. 
It is interesting to clarify whether the theory corresponding to this data has interesting properties as other theories mentioned above.

\vspace{1cm}

\section{Theories $\Gamma$ with two nodes}  \label{2node}

\vspace{5mm}

In this section, we discuss a simple class of theories in ${\cal C}_0$ whose diagrams contain only two nodes, that is, $n_g=2$. 
Obviously, the corresponding diagrams are bipartite, and therefore, the saddle-point equations (\ref{SP_u}) can be written as (\ref{SP_bipartite}). 
The equations (\ref{SP_bipartite}) can be reduced to the homogeneous ones (\ref{SP_homogeneous}) if and only if $n:=n^{12}$ does not satisfy 
\begin{equation}
4-n^2\ =\ 0. 
\end{equation}
The degenerate theories with $n_g=2$ include, for example, ABJM theory \cite{Aharony:2008ug}, ABJ theory \cite{Aharony:2008gk}, 
ABJM theory with flavors \cite{Santamaria:2010dm} and GT theory \cite{Gaiotto:2009mv}. 
The planar solution of ABJ(M) theory was obtained in \cite{Marino:2009jd}, 
flavored ABJM theory was solved in \cite{Santamaria:2010dm} and GT theory was investigated in \cite{Suyama:2010hr}. 
In the following, we will focus on the remaining cases $n\ne2$. 
The technique for solving (\ref{SP_homogeneous}) in this section is a generalization of the one developed in \cite{Eynard:1995nv}\cite{Eynard:1995zv}. 

As explained in the previous section, the study of this class also gives us the planar solution of any theory $\Gamma$ with a single node to which 
various edges 
corresponding to {\bf f}, {\bf ad}, {\bf s} and {\bf as} are connected. 

\vspace{5mm}

The equations (\ref{SP_homogeneous}) for $n_g=2$ are 
\begin{eqnarray}
\omega^1(y_1^+)+\omega^1(y_1^-)-n\,\omega^2(y_1) &=& 0,  
   \label{2node_homogeneous1} \\
\omega^2(y_2^+)+\omega^2(y_2^-)-n\,\omega^1(y_2) &=& 0,  
   \label{2node_homogeneous2}
\end{eqnarray}
where $y_1\in[p_1,q_1]$, $y_2\in[p_2,q_2]$. 
For a while, the dependence on $\xi$ will be ignored. 
The configuration of these intervals can be arbitrary in the complex plane $\mathbb{C}$. 
In fact, it is enough to discuss the case where the intervals are symmetric, that is, 
\begin{equation}
p_1\ =\ -q_2, \hspace{5mm} q_1\ =\ -p_2. 
\end{equation}
Indeed, there is an ${\rm SL}(2,\mathbb{C})$ transformation which maps the four points $(p_1,q_1,p_2,q_2)$ to $(-q,-p,p,q)$. 
Explicitly, it is given as $\varphi_1\circ\varphi_2$ where 
\begin{equation}
\varphi_1(z)\ :=\ c\frac{z+\sqrt{a}}{z-\sqrt{a}}, \hspace{5mm} \varphi_2(z)\ :=\ \frac{q_1-q_2}{q_1-p_1}\frac{z-p_1}{z-q_2}, 
\end{equation}
and 
\begin{equation}
a\ :=\ \varphi_2(p_2), \hspace{5mm} c^2\ :=\ \frac{\sqrt{a}-1}{\sqrt{a}+1}. 
\end{equation}
The constant $c$ is chosen such that 
\begin{equation}
pq\ =\ 1 
   \label{2node_sym}
\end{equation}
is satisfied. 
In the following, the intervals are assumed to be symmetric, satisfying the condition (\ref{2node_sym}). 
The solution for a generic configuration can be obtained through the pull-back by $\varphi_1\circ\varphi_2$. 

By definition, $\omega^1(z)$ and $\omega^2(z)$ do not have a branch cut on $[p,q]$ and $[-q,-p]$, respectively. 
This is expressed as 
\begin{equation}
\omega^1(y_2^+)\ =\ \omega^1(y_2^-), \hspace{5mm} \omega^2(y_1^+)\ =\ \omega^2(y_1^-). 
   \label{2node_no-cut}
\end{equation}
It is convenient to introduce a vector-valued function 
\begin{equation}
\omega(z)\ :=\ \left( \omega^1(z),\omega^2(z) \right). 
\end{equation}
In terms of $\omega(z)$, the equations (\ref{2node_homogeneous1})(\ref{2node_homogeneous2})(\ref{2node_no-cut}) can be written as 
\begin{equation}
\omega(y_1^+)\ =\ \omega(y_1^-)M_1, \hspace{5mm} \omega(y_2^+)\ =\ \omega(y_2^-)M_2,  
   \label{2node_matrix}
\end{equation}
where 
\begin{eqnarray}
M_1\ =\ \left[ 
\begin{array}{cc}
-1 & 0 \\
n & 1
\end{array}
\right], \hspace{5mm} M_2\ =\ \left[ 
\begin{array}{cc}
1 & n \\
0 & -1
\end{array}
\right]. 
   \label{monodromy matrix}
\end{eqnarray}
Note that these matrices satisfy 
\begin{equation}
M_1^2\ =\ M_2^2\ = \ I, 
\end{equation}
where $I$ is the unit matrix.

\vspace{5mm}

\subsection{General solution}   \label{general solution}

\vspace{5mm}

\begin{figure}[tbp]
\begin{center}
\includegraphics{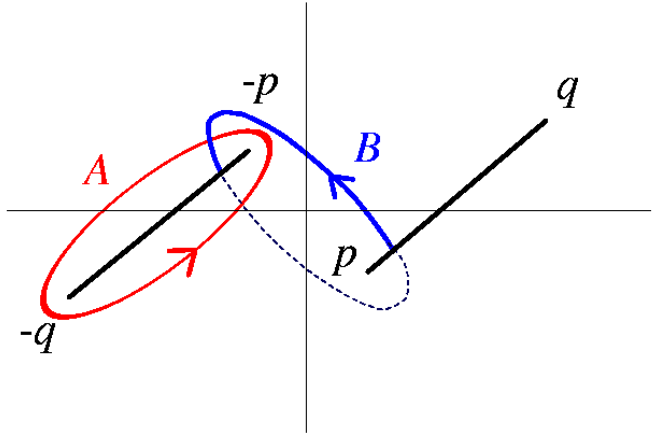}
\end{center}
\caption{
The red curve is the $A$-cycle, and the blue curve which passes through branch cuts is the $B$-cycles. 
}
\label{cycles}
\end{figure}

As in \cite{Suyama:2012uu}, it is convenient to change the variable. 
The new variable $u$ has a geometric meaning \cite{BorotEynard} of an elliptic curve $E$ defined by 
\begin{equation}
y^2\ =\ (x^2-p^2)(x^2-q^2). 
\end{equation}
Let $\lambda$ be the holomorphic 1-form on $E$ normalized such that 
\begin{equation}
\int_A\lambda\ =\ -\tau, \hspace{5mm} \int_B\lambda\ =\ 1
\end{equation}
with $\mbox{Im}(\tau)>0$. 
The $A$-cycle and the $B$-cycle are depicted in Figure \ref{cycles}. 
This unconventional choice is due to the fact that we are interested in the large $|\tau|$ limit in the above definition 
where the 't~Hooft couplings are expected to diverge. 
In the usual convention, this limit correspond to $\tau\to0$, and we have performed the $S$-transformation from the beginning. 
In the opposite limit where the branch cuts shrink to points, one may perform the $S$-transformation, or 
a perturbative calculation developed for ABJM theory in \cite{Suyama:2009pd} can be adapted to 
provide a perturbative solution. 

The coordinate transformation to the new variable $u$ is defined by the Abel-Jacobi map 
\begin{equation}
u(z)\ := \ \int_{p}^z\lambda
\end{equation}
with the base point $p$. 
The image of the whole $z$-plane is a parallelogram 
\begin{equation}
E_u\ :=\ \left\{ \,\alpha+\beta\tau\in\mathbb{C}\ \Big|\ 0\le\alpha\le\frac12, |\beta|\le\frac12 \right\}. 
\end{equation}
The horizontal edges of $E_u$ are identified. 
Roughly speaking, the left edge corresponds to the union of the upper- and the lower-side of the branch cut $[p,q]$, and the right edge corresponds to 
$[-q,-p]$. 
See Figure \ref{AJ_map}. 
Some properties of the map are shown in Appendix \ref{Abel-Jacobi}. 

\begin{figure}[tbp]
\begin{center}
\includegraphics{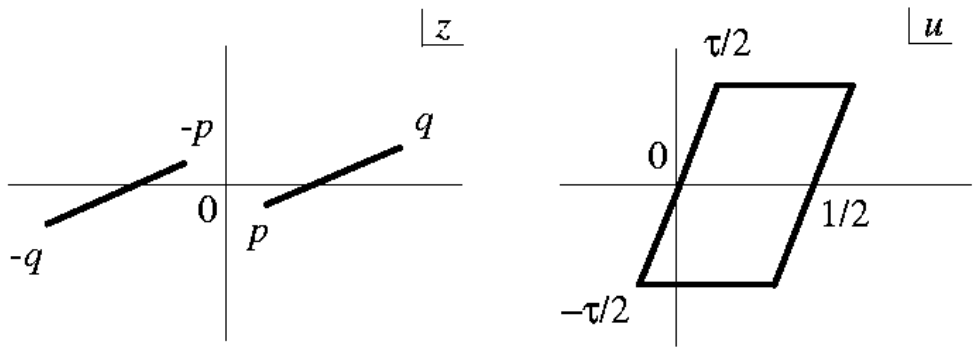}
\end{center}
\caption{
The Abel-Jacobi map maps the whole $z$-plane with branch cuts into the parallelogram in the $u$-plane. 
The upper- and the lower-edges of the parallelogram are identified. 
The upper-half and the lower half of the left edge are the images of the same curve on $\mathbb{C}$, lying on the different Riemann sheets. 
Similar for the right edge. 
}
\label{AJ_map}
\end{figure}
\vspace{5mm}

One can show that, if a vector-valued function $\Omega(u)$ defined on the $u$-plane satisfies the following conditions 
\begin{equation}
\Omega(u+1)\ =\ \Omega(u)M_2M_1,\hspace{5mm} \Omega(u+\tau)\ =\ \Omega(u), \hspace{5mm} \Omega(-u)\ =\ \Omega(u)M_2, 
   \label{2node_periodic}
\end{equation}
for any $u\in\mathbb{C}$, then $\omega(z)$ given by the pull-back, 
\begin{equation}
\omega(z)\ =\ \Omega(u(z)), 
\end{equation}
is a solution of (\ref{2node_matrix}). 

\vspace{5mm}

The equations (\ref{2node_periodic}) can be solved easily by diagonalizing the matrix $M_2M_1$. 
This can be done by a similarity transformation defined by a matrix 
\begin{equation}
S\ :=\ \left[
\begin{array}{cc}
e^{\frac12\pi i\nu} & e^{-\frac12\pi i\nu} \\
e^{-\frac12\pi i\nu} & e^{\frac12\pi i\nu}
\end{array}
\right], 
\end{equation}
which is non-singular for $n\ne2$. 
Here $\nu$ is related to $n$ as 
\begin{equation}
n\ =\ -2\cos\pi\nu. 
   \label{nu}
\end{equation}
The transformed function $\Omega_S(u):=\Omega(u)S$ satisfies, in terms of components, 
\begin{eqnarray}
& \Omega_S^1(u+1)\ =\ e^{2\pi i\nu}\Omega_S^1(u), \hspace{5mm} \Omega_S^1(u+\tau)\ =\ \Omega_S^1(u), & 
   \label{2node_comp1} \\
& \Omega_S^2(u)\ =\ -e^{\pi i\nu}\Omega_S^1(-u). &
   \label{2node_comp2}
\end{eqnarray}

Using the identities of the theta functions, one can show that the following function 
\begin{equation}
G(u)\ :=\ e^{2\pi i\nu u}\frac{\vartheta_1(u-u_0+\frac\nu2\tau,\tau)\vartheta_1(u-u_0+\frac12+\frac\nu2\tau,\tau)}
 {\vartheta_1(u-u_0,\tau)\vartheta_1(u-u_0+\frac12,\tau)}, 
\end{equation}
where $u_0\in\mathbb{C}$ can be arbitrary, satisfies the equations (\ref{2node_comp1}). 
Suppose that $\Omega_S^1(u)$ is given as 
\begin{equation}
\Omega_S^1(u)\ =\ R(u)G(u). 
   \label{2node_general}
\end{equation}
Then, the equations (\ref{2node_comp1}) imply that $R(u)$ is an elliptic function on $E$. 
$\Omega_S^2(u)$ is given in terms of $R(u)$ and $G(u)$ by (\ref{2node_comp2}). 
Explicitly, 
\begin{equation}
\Omega_S^2(u)\ =\ -R(-u)G\left( -u+\frac12 \right). 
\end{equation}
Note that there is a simpler choice for $G(u)$. 
A different $G(u)$ just results in a different $R(u)$. 

\vspace{5mm}

Now we recover the $\xi$-dependence. 
It is obvious that the dependence of $\omega(z,\xi)$ on $\xi$ is encoded in the elliptic function $R(u,\xi)$ only. 
The functional form of $R(u,\xi)$ is constrained since $\omega(z,\xi)$ must have the appropriate pole structure. 

For simplicity, we consider the case $n^a=0$ first. 
The general case will be discussed in subsection \ref{matter}. 
In the case $n_g=2$, the vector-valued function $r(z,\xi):=(r^1(z,\xi),r^2(z,\xi))$, whose components were introduced 
in (\ref{rational part}), is given in terms of $f(z,\xi):=(f^1(z,\xi),f^2(z,\xi))$ as 
\begin{equation}
r(z,\xi)\ =\ \frac1{4-n^2}f(z,\xi)\left[
\begin{array}{cc}
2 & n \\
n & 2 
\end{array}
\right]. 
\end{equation}
The function $r(z,\xi)$ has poles at $z=\pm\xi$ with the residues 
\begin{eqnarray}
\mbox{Res}_{-\xi}r(z,\xi) &=& \frac{2\kappa^1}{4-n^2}(2,n), \\
\mbox{Res}_{+\xi}r(z,\xi) &=& -\frac{2\kappa^2}{4-n^2}(n,2). 
\end{eqnarray}
Therefore, $\omega(z,\xi)=\Omega(u(z),\xi)$ must have poles with residues which have opposite signs to the above ones. 
In terms of $\omega_S(z,\xi):=\Omega_S(u(z),\xi)$, the required residues are 
\begin{eqnarray}
\mbox{Res}_{-\xi}\omega_S(z,\xi) &=& -\frac{i\kappa_1}{\sin\pi\nu}\left( e^{-\frac12\pi i\nu},-e^{\frac12\pi i\nu} \right), 
   \label{residue1} \\
\mbox{Res}_{+\xi}\omega_S(z,\xi) &=& -\frac{i\kappa_2}{\sin\pi\nu}\left( e^{\frac12\pi i\nu},-e^{-\frac12\pi i\nu} \right). 
   \label{residue2}
\end{eqnarray}

In the following, we abbreviate $R(u(z),\xi)$ as $R(z,\xi)$ etc. 
Since $\omega_S(z,\xi)$ must have poles at $z=\pm\xi$, $R(z,\xi)$ can be written as 
\begin{equation}
R(z,\xi)\ =\ \frac1{z^2-\xi^2}R_0(z,\xi)
\end{equation}
with another elliptic function $R_0(z,\xi)$ on $E$. 
To write down the most general form of $R_0(z,\xi)$ explicitly, it is convenient to choose $u_0$ in $G(z)$ as 
\begin{equation}
u_0\ =\ u_\infty\ :=\ u(\infty). 
\end{equation}
For this choice, the function 
\begin{equation}
F(z,\xi) \ :=\ \frac1{z^2-\xi^2}G(z)
\end{equation}
has simple poles at $z=\pm\xi$, simple zeros at $z=z_\nu,\infty,\widetilde{-z_\nu},\widetilde{\infty}$, where the tilded points lie on the second 
Riemann sheet. 
Here $z_\nu$ is defined such that 
\begin{equation}
u(z_\nu) \ =\ u_\infty-\frac\nu2\tau. 
\end{equation}
The elliptic function $R_0(z,\xi)$ may have poles which can be canceled by the zeros of $F(z,\xi)$. 
In terms of the divisor, 
\begin{equation}
(R_0(z))\ \ge\ -(z_\nu)-(\infty)-(\widetilde{-z_\nu})-(\widetilde{\infty}). 
\end{equation}
Recall that $v(z,\xi)$ is assumed to be finite at branch points. 

The Riemann-Roch theorem implies that there are four such functions linearly independent on $\mathbb{C}$. 
Therefore, $R_0(z,\xi)$ can be written as 
\begin{equation}
R_0(z,\xi)\ =\ \sum_{i=1}^4c_i(\xi)f_i(z)
\end{equation}
for some basis functions $f_i(z)$. 

In summary, it has been shown that the general form of $\omega_S(z,\xi)$ is 
\begin{eqnarray}
\omega_S^1(z,\xi) &=& \frac1{z^2-\xi^2}G(z)\left[ c_1(\xi)f_1(z)+c_2(\xi)f_2(z)+c_3(\xi)f_3(z)+c_4(\xi)f_4(z) \right], \\
\omega_S^2(z,\xi) &=& -\frac1{z^2-\xi^2}G(-z)
   \left[ c_1(\xi)f_1(\widetilde{z})+c_2(\xi)f_2(\widetilde{z})+c_3(\xi)f_3(\widetilde{z})+c_4(\xi)f_4(\widetilde{z}) \right]. 
\end{eqnarray}
The four parameters $c_i(\xi)$ can be determined by the four conditions for the residues, two for $\omega_S^1(z,\xi)$ and two for $\omega_S^2(z,\xi)$. 
Therefore, the vector-valued resolvent $v(z)$ obtained from $\omega_S(z,\xi)$ can be completely determined by the data given so far. 

\vspace{5mm}

It is important to note that the above analysis can be used to show that the (vector-valued) 
solution $v(z)$ of the saddle-point equations (\ref{SP_bipartite}) is unique. 
Suppose that there are two solutions $v(z)$ and $v'(z)$. 
The difference $\delta v(z):=v(z)-v'(z)$ is a solution of the homogeneous equations (\ref{SP_homogeneous}). 
Therefore, the function $\delta v^1_S(z)$, which is the first component of $\delta v(z)S$, can be written as 
\begin{equation}
\delta v^1_S(z)\ =\ R(z)G(z). 
\end{equation}
In this case, $R(z)$ does not have poles at $z=\pm\xi$. 
It must have zeros at $z=\infty,\widetilde{\infty}$ to cancel the poles of $G(z)$. 
Since $R(z)$ is an elliptic function, there must be at least two poles. 
However, the only allowed positions of the poles of $R(z)$ are at $z=z_\nu,\widetilde{-z_\nu}$ since $\delta v_S(z)$ should not have any poles. 
The divisor of $R(z)$ is therefore 
\begin{equation}
(R(z))\ =\ -(z_\nu)-(\widetilde{-z_\nu})+(\infty)+(\widetilde{\infty}). 
\end{equation}
Another constraint for elliptic functions is that the $u$-coordinates of these zeros and poles must satisfy 
\begin{equation}
-u(z_\nu)-u(\widetilde{-z_\nu})+u(\infty)+u(\widetilde{\infty})\ \in\ \Lambda, 
\end{equation}
where 
\begin{equation}
\Lambda\ :=\ \mathbb{Z}+\mathbb{Z}\tau. 
\end{equation}
This then implies 
\begin{equation}
\nu\tau\ \in\ \Lambda. 
   \label{2node_constraint}
\end{equation}
Therefore, for generic values of $\nu$ and $\tau$, the only allowed solution is $\delta v(z)=0$, implying the uniqueness of the solution of 
(\ref{SP_bipartite}). 
It will be shown below that if (\ref{2node_constraint}) is satisfied, then there is actually no solution. 

\vspace{5mm}

\subsection{The solution in detail}

\vspace{5mm}

It is necessary to choose the basis functions $\{f_i(z)\}$ for the elliptic functions in order to further investigate the properties of the solution. 
A simple choice of the basis is 
\begin{equation}
f_1(z)\ =\ 1, \hspace{5mm} f_2(z)\ =\ z, \hspace{5mm} f_3(z)\ =\ \frac{y(z)+y(z_\nu)}{z-z_\nu}, \hspace{5mm} f_4(z)\ =\ \frac{y(z)-y(z_\nu)}{z+z_\nu}, 
   \label{elliptic basis}
\end{equation}
where 
\begin{equation}
y(z)\ :=\ \sqrt{(z^2-p^2)(z^2-q^2)}. 
\end{equation}
Apparently, these are linearly independent on $\mathbb{C}$ since they have different pole structures. 
The coefficients $\{c_i(\xi)\}$ are determined by the conditions (\ref{residue1})(\ref{residue2}) which can be written as 
a linear equation 
\begin{equation}
M\cdot c\ =\ \rho, 
   \label{linear}
\end{equation}
where $c$ and $\rho$ are 4-component vectors which are given by $c_i(\xi)$ and the residues, respectively, 
and $M$ is a $4\times4$ matrix constructed in terms of the values $f_i(\pm\xi)$. 
The determinant of $M$ is 
\begin{equation}
\mbox{det}\,M\ =\ \frac{32z_\nu\xi^2y(\xi)^2}{(\xi^2-z_\nu^2)^2}. 
\end{equation}
For a generic $\xi$, this is non-vanishing if and only if $z_\nu$ is non-zero and finite. 
In this case, $\{c_i(\xi)\}$ are determined uniquely. 
The explicit expression of $\{c_i(\xi)\}$ is given in Appendix \ref{coefficients}. 

The linear equation (\ref{linear}) becomes singular, and therefore there is no solution, if $z_\nu=0,\infty$. 
This condition can be written as 
\begin{equation}
\nu\tau\ \in\ \Lambda, 
   \label{2node_no solution}
\end{equation}
which was encountered as the condition for which the uniqueness of the solution may fail. 

Recall that the parameter $\nu$ was defined as (\ref{nu}). 
To realize a real $n$, $\nu$ must take the following values, 
\begin{equation}
\nu\ \in \ \left\{ 
\begin{array}{cc}
(\frac12,1] & (0<n\le2), \\
1+i\mathbb{R} & (n>2). 
\end{array}
\right. 
\end{equation}
The sign of $\mbox{Im}(\nu)$ for $n>2$ is irrelevant due to the uniqueness of the solution. 

The singularity at $\nu=1$ ($n=2$) is anticipated since 
this is when the theory becomes degenerate. 
This singularity is special compared with the other ones since it is singular for any value of $\tau$. 
One might expect that a limit $\nu\to1$ would be well-defined. 
In fact, a regular behavior in the 't~Hooft coupling in this limit was observed in \cite{Suyama:2012uu}. 
However, this is not the case at the level of the resolvents, as one can check using the explicit form of the coefficients shown in Appendix 
\ref{coefficients}. 
The individual coefficients are apparently divergent in the limit $\nu\to1$. 
At the level of the 't~Hooft couplings, there is a possibility of a cancellation of singularities among the coefficients. 
Since $f_i(z)$ are linearly independent, the resolvent is not well-defined in the limit. 

If $n>2$, then there are many values of $\tau$ for which the solution does not exist. 
Since the solution is well-defined for a generic value of $\tau$, the theory itself should make sense, and these singularities must be meaningful. 
In fact, these singularities appear as divergences of the 't~Hooft couplings, as will be shown in subsection \ref{coupling}. 
Another viewpoint on these singularities will be given in section \ref{reform}. 

\vspace{5mm}

It has turned out that the solution of (\ref{2node_comp1}) is uniquely determined. 
In fact, this also implies that the resulting solution $\omega(z,\xi)$ has the following inversion property, 
\begin{equation}
\omega(z^{-1},\xi)\ =\ -\xi^{-2}\omega(z,\xi^{-1}), 
   \label{2node_inversion}
\end{equation}
which, combined with 
\begin{equation}
f^a(z^{-1},\xi)\ =\ -\xi^{-2}f^a(z,\xi^{-1}), 
\end{equation}
implies the inversion property (\ref{inversion}). 
To show this, note that $u(z)$ satisfies 
\begin{equation}
u(z^{-1})\ =\ u(z)+\frac\tau2. 
\end{equation}
See Appendix \ref{Abel-Jacobi}. 
This implies $G(z^{-1})=G(u(z)+\frac\tau2)$. 
Since $G(u+\frac\tau2)$ also satisfies (\ref{2node_comp1}), this can be written as a product of $G(u)$ with an elliptic function $R'(u)$. 
Due to the same argument, one finds that $\omega^1_S(z^{-1},\xi)$ is given as 
\begin{equation}
\omega^1_S(z^{-1},\xi)\ =\ \frac1{z^2-\xi^{-2}}G(z)\left[ c'_1(\xi)f_1(z)+c'_2(\xi)f_2(z)+c'_3(\xi)f_3(z)+c'_4(\xi)f_4(z) \right]. 
\end{equation}
Noticing a trivial identity 
\begin{equation}
\frac1{z^{-1}-\xi}\ =\ -\frac1{\xi^2}\frac1{z-\xi^{-1}}-\frac1\xi, 
\end{equation}
one finds that the residues of $\omega(z^{-1},\xi)$ at $z=\pm\xi^{-1}$ are related to the residues of $\omega(z,\xi)$ as 
\begin{equation}
\mbox{Res}_{\pm\xi^{-1}}\omega(z^{-1},\xi)\ =\ -\xi^{-2}\mbox{Res}_{\pm\xi}\omega(z,\xi). 
\end{equation}
This implies that the coefficients $\{c_i'(\xi)\}$ are related to $\{c_i(\xi)\}$ as 
\begin{equation}
c_i'(\xi)\ =\ -\xi^{-2}c_i(\xi^{-1}), 
\end{equation}
which then implies (\ref{2node_inversion}).

\vspace{5mm}

\subsection{Chern-Simons-adjoint theory revisited}

\vspace{5mm}

In \cite{Suyama:2012uu}, ${\cal N}=3$ Chern-Simons-adjoint theories were discussed. 
These are theories in ${\cal C}$ a member of which corresponds to a diagram $\Gamma_{\bf ad}$ 
consisting of only a single node with an arbitrary number of {\bf ad} edges. 
As explained in subsection \ref{equivalence}, the solution in \cite{Suyama:2012uu} 
can be reproduced from the solution obtained in this section by imposing a suitable constraint. 

We have considered a theory $\Gamma$ consisting of two nodes and a number of {\bf bf} edges. 
The constraint requires that the eigenvalue distributions of two nodes of $\Gamma$ are the same, making {\bf bf} become equivalent to 
half the number of {\bf ad}. 
In terms of the resolvent, the constraint is 
\begin{equation}
v^1(z)\ =\ v^2(-z). 
\end{equation}
This implies that the vector-valued function $\omega(z,\xi)$ must satisfy 
\begin{equation}
\omega(z,\xi)\ =\ \omega(-z,\xi)\left[ 
\begin{array}{cc}
0 & 1 \\
1 & 0
\end{array}
\right]. 
\end{equation}
This is realized by the pull-back of $\Omega(u,\xi)$ satisfying 
\begin{equation}
\Omega^1\left( -u+\frac12 \right)\ =\ \Omega^2(u). 
\end{equation}
Using (\ref{2node_comp2}), this can be written as 
\begin{equation}
\Omega^1_S\left( u+\frac12 \right)\ =\ -e^{\pi i\nu}\Omega^1_S(u). 
\end{equation}
If $\Omega^1_S(u)$ satisfies this equation and the second equation in (\ref{2node_comp1}), then the remaining equation is automatically satisfied. 
$\Omega^2_S(u)$ is then determined by the equation (\ref{2node_comp2}). 

The function $G(u)$ satisfies 
\begin{equation}
G\left( u+\frac12 \right) \ =\ e^{\pi i\nu}G(u). 
\end{equation}
If $\Omega^1_S(u)$ is written as in (\ref{2node_general}), then 
one finds that $R(u)$ is not just an elliptic function but it satisfies 
\begin{equation}
R\left( u+\frac12 \right) \ =\ -R(u). 
\end{equation}
As a function of $z$, this implies 
\begin{eqnarray}
& & c_1(\xi)-c_2(\xi)z+c_3(\xi)f_4(z)+c_4(\xi)f_3(z) \nonumber \\ 
&=& -\left[ c_1(\xi)+c_2(\xi)z+c_3(\xi)f_3(z)+c_4(\xi)f_4(z) \right]. 
\end{eqnarray}
Therefore, the solution of a Chern-Simons-adjoint theory is written by the coefficients $\{c_i(\xi)\}$ satisfying 
\begin{equation}
c_1(\xi)\ =\ 0, \hspace{5mm} c_3(\xi)\ =\ -c_4(\xi). 
   \label{constraint-CSA}
\end{equation}
The remaining coefficients are determined by requiring that $\omega^1(z,\xi)$ has the appropriate pole structure. 
Since $\omega^2(z,\xi)$ is now given by $\omega^1(z,\xi)$, there are only two conditions. 

For the Chern-Simons-adjoint theory, one has to have $\kappa_1=\kappa_2$ since there is only one ${\rm U}(N)$ gauge group factor. 
The explicit formulas for the coefficients shown in Appendix \ref{coefficients} show that this is compatible with (\ref{constraint-CSA}). 
Note that the explicit functional form given in \cite{Suyama:2012uu} is different from the one obtained in this paper. 
This is simply because the choice of $G(z)$ in this paper is different from the one in \cite{Suyama:2012uu}. 
Since the solution is proved to be unique, these two expressions define the same function.

\vspace{5mm}

\subsection{'t~Hooft couplings}  \label{coupling}

\vspace{5mm}

Recall that the 't~Hooft couplings are given in terms of resolvents as 
\begin{equation}
t^a\ =\ -v^a(0). 
\end{equation}
This can be written in terms of $\omega(z,\xi)$ as in (\ref{'t Hooft}). 
It is more convenient to consider the transformed coupling $t_S:=(t^1,t^2)S$. 
For example, $t^1_S$ is given as 
\begin{equation}
t^1_S\ =\ -\int_Cd\xi\left[ \left( \frac{\kappa^1+\kappa^2}{\cos\frac{\pi\nu}2}+i\frac{\kappa^1-\kappa^2}{\sin\frac{\pi\nu}2} \right)\frac1{\xi(1-\xi)}
 +\Omega_S(0,\xi) \right], 
\end{equation}
where 
\begin{equation}
\Omega_S^1(0,\xi)\ =\ -\frac1{\xi^2}G(0)\left[ c_1(\xi)-\frac{y(z_\nu)}{z_\nu}(c_3(\xi)+c_4(\xi))+\frac1{z_\nu}\left( c_3(\xi)-c_4(\xi) \right) \right]. 
\end{equation}
One can check that the integral is well-defined. 
Indeed, the integrand behaves as $\xi^{-2}$ in the limit $\xi\to\infty$, while it is finite in the limit $\xi\to 0$. 

The 't~Hooft coupling $t_S$ depends on the positions $p,q$ of the branch points. 
In fact, since $pq=1$ is assumed, $t_S$ is a vector-valued function of $p$, say, and the domain of $t_S(p)$ is the unit disk $D$. 
The behavior of $t_S(p)$ for a generic $p$ would be quite complicated. 
We are interested in the possible divergent behavior of $t_S(p)$. 
In the following, we investigate not just the positions where $t_S(p)$ diverges but also how it diverges. 

In ABJM theory and other Chern-Simons-matter theories with known gravity duals, interesting behaviors are observed when the 't~Hooft couplings 
diverge. 
Recall that imaginary values of $\{t^a\}$ correspond to the physical values for the original Chern-Simons-matter theories. 
Therefore, a divergent behavior of $t_S(p)$ would be interesting in the context of the original Chern-Simons-matter theories 
if both $t^1(p),t^2(p)$ have divergent imaginary parts while the real parts could be tuned to vanish in the limit. 

\vspace{5mm}

Apparently, $t_S(p)$ diverges when $z_\nu=0,\infty$. 
These are simple poles of $t_S(p)$ in $D$. 
It can be shown that the condition for $z_\nu=0$ is 
\begin{equation}
\nu\tau\ \in\ \Lambda. 
\end{equation}
Obviously, there is no solution for $0<n<2$ since for this range $\frac12<\nu<1$. 
For the case $n>2$, there may be many solutions. 
In fact, if $\tau$ is purely imaginary, which is the case when $p$ is real, then there are infinitely many solutions. 
This implies that $t_S(p)$ has infinitely many simple poles on the real axis. 

The fact that there are infinitely many simple poles for $t_S(p)$ implies that there are infinitely many values of $p$ for which 
$t_S(p)$ may have a prescribed large value. 
That is, for a given value of $t_S$, there are infinitely many saddle-points for the original partition function (\ref{localizedZ}). 
In this case, one has to choose one of the saddle-points having the smallest free energy. 
We will not discuss further about the criterion for choosing the dominant saddle-point. 
Naively, it could be expected that the minimum of the free energy is of order $k^2$ (recall that $N_a\propto k$ in our planar limit) 
which would come from a saddle point corresponding to a value of $p$ away 
from the origin. 
Similar results were found in \cite{Marino:2012az}. 

\vspace{5mm}

The other possibilities for the divergence of $t_S(p)$ are also simple. 
Since $t_S(p)$ is given in terms of contour integrals, divergences appear when a singularity of the integrand approaches one of the endpoints of the 
contour, or when the contour is pinched by two singularities of the integrand. 

An endpoint singularity appears when the branch point at $z=p$ in $R_0(z)$ approaches the origin. 
Note that the other endpoint, at infinity, does not provide any divergence, and 
$G(z)$ does not have singularities which can approach the origin. 
If $p$ is very close to the origin, then a large contribution comes from the term 
\begin{equation}
\frac{\kappa^1+\kappa^2}{2\cos\frac{\pi\nu}2}\frac1\xi\left[ \frac1{y(\xi)}+1 \right]
\end{equation}
in the integrand. 
The integral of this term gives a contribution 
\begin{equation}
t^1_S(p)\ \sim -\frac{\kappa^1+\kappa^2}{2\cos\frac{\pi\nu}2}\log p
\end{equation}
which is divergent in the limit $p\to0$. 
It turns out that $t^2_S(p)$ also diverges in the same way. 
The original couplings $\{t^a\}$ then behave as 
\begin{equation}
t^1,t^2\ \sim \ -\frac{\kappa^1+\kappa^2}{4\cos^2\frac{\pi\nu}{2}}\log p. 
\end{equation}
The divergent part of the right-hand side is always real. 
To obtain large imaginary values for $t^a$, one has to continue $p$ along a path which winds around the origin many times, as was found in 
\cite{Suyama:2012uu} for Chern-Simons-adjoint theories. 
The vanishing real part will be ensured only if $|p|=O(1)$. 

Note that the limit $p\to0$ above is assumed not to be taken along the real axis for the cases $n>2$, since there are poles. 
If one takes a limit such that 
\begin{equation}
\tau\ =\ \tau_1+i\tau_2, \hspace{5mm} |\tau_1|\ =\ c\ <\ \frac12, \hspace{5mm} \tau_2\to+\infty 
\end{equation}
for a constant $c$, then the distance between $\nu\tau$ and the lattice points in $\Lambda$ is bounded from below, and therefore $z_\nu^{-1}$ can be 
regarded as a smooth function of $p$. 

The pinch singularity may appear when $p$ approaches either $\pm i$ or $-1$. 
In any cases, the resulting branch cuts, and therefore the eigenvalue distributions, have finite lengths. 
This is an uninteresting situation in the following sense. 
In ABJM theory, the expectation value of the BPS Wilson loop grows exponentially as the 't~Hooft coupling grows. 
This is quite naturally explained from the dual string point of view \cite{Rey:1998ik}\cite{Maldacena:1998im}\cite{Rey:1998bq}. 
On the other hand, if a Chern-Simons-matter theory has a finite size branch cut, then the expectation value 
of the BPS Wilson loop corresponding to the cut must have a bounded value. 
The expectation value of a Wilson loop, say the one for ${\rm U}(N_a)$ gauge group factor, is given as 
\begin{equation}
\langle W^a \rangle\ =\ \frac1{N_a}\sum_{i_a=1}^{N_a}\bar{z}^a_{i_a}  
\end{equation}
where $\{\bar{z}^a_{i_a}\}$ are the saddle-point values. 
The absolute value of this is bounded as 
\begin{equation}
|\langle W^a \rangle|\ \le\ |q| 
\end{equation}
since $\pm q$ are the eigenvalues with the largest absolute value. 
Therefore, if the branch cuts are of finite size even in the large 't~Hooft coupling limit, then the magnitudes of the 
quantities $\langle W^a\rangle$ are bounded in the limit. 
It seems to be difficult to explain this behavior by a smooth worldsheet in a classical geometry. 

\vspace{5mm}

In summary, the divergent behavior of $\{t^a\}$ as functions of $p$ has a rather simple pattern. 
One might wonder whether the properties for the theories with $n_g=2$ found so far would be generic ones for the theories in ${\cal C}_0$. 
We will investigate a generic non-degenerate 
theory in ${\cal C}_0$ in section \ref{reform} and again find a similar behavior of the 't~Hooft couplings and the Wilson loops. 
It might suggest that any non-degenerate theory in ${\cal C}_0$ 
would not have simple gravity duals with the standard dictionary between the field theory and the gravity.

\vspace{5mm}

\subsection{Adding fundamental matters}  \label{matter}

\vspace{5mm}

Now, let us consider the effect of adding {\bf f} to the theory discussed so far. 
The homogeneous equations (\ref{SP_homogeneous}) are the same, and therefore the functions $\{\omega^a_S(z,\xi)\}$ can be written as 
\begin{equation}
\omega^1_S(z,\xi)\ =\ R(u(z))G(u(z)), \hspace{5mm} \omega^2_S(z,\xi)\ =\ -R(-u(z))G\left( -u(z)+\frac12 \right). 
\end{equation}
The difference from the analysis in the previous subsections appears in the form of $R(u)$. 
Since the matters {\bf f} add poles at $z=\pm1$ to $v^a_1(z)$ in (\ref{fund}), the elliptic function $R(u(z))$ should have the following form 
\begin{equation}
R(u(z))\ =\ \frac1{(z^2-1)(z^2-\xi^2)}R_1(u(z)) 
\end{equation}
with another elliptic function $R_1(u)$. 
It turns out that the divisor of $R_1(u)$ is constrained as 
\begin{equation}
(R_1(u))\ \ge -(z_\nu)-(\widetilde{-z_\nu})-3(\infty)-3(\widetilde{\infty}). 
\end{equation}
The Riemann-Roch theorem implies that such functions form an eight-dimensional linear space. 
The residues of the extra poles at $z=\pm1$ provide four more conditions. 
Therefore, the solution is again completely determined by the data given so far. 

The basis functions for $R_1(z)$ are $f_1(z),\cdots,f_4(z)$ given in (\ref{elliptic basis}) and 
\begin{equation}
f_5(z)\ =\ z^2, \hspace{5mm} f_6(z)\ =\ y(z), \hspace{5mm} f_7(z)\ =\ z^3, \hspace{5mm} f_8(z)\ =\ z\,y(z). 
\end{equation}
The existence of the solution depends on the determinant of an $8\times8$ matrix $M$ whose components are the values of the basis functions 
at the poles. 
The determinant is 
\begin{equation}
\mbox{det}\,M\ =\ -\frac{512z_\nu y(1)^2\xi^2(\xi^2-1)^4y(\xi)^2}{(z_\nu^2-1)^2(\xi^2-z_\nu^2)^2}. 
\end{equation}
New singularities would appear when $z_\nu=\pm1$. 
Again, this singularity does not appear when $0<n<2$, while it would appear for some particular values of $\tau$ when $n>2$. 

The addition of {\bf f} introduces two poles at $z=\pm1$ to $\omega_S(z,\xi)$. 
The functional form of $\omega_S(z,\xi)$ is different from the one in the absence of {\bf f}. 
However, the behavior of $t^a(p)$ would not be drastically modified by the addition of {\bf f} since $t^a$ is determined by the values 
$\omega^a_S(z,\xi)$ at $z=0$ which is away from the new poles. 
Note that, although not directly related to our cases, in the case of the flavored ABJM theory \cite{Santamaria:2010dm} 
the effects of {\bf f} appear, for example, in a modified coefficient in the exponent of 
the expectation value $\langle W \rangle$ of the BPS Wilson loop. 
However, the qualitative properties of $\langle W \rangle$ as a function of the 't~Hooft coupling are kept intact.

\vspace{1cm}

\section{Non-degenerate theory $\Gamma$ in ${\cal C}_0$}  \label{reform}

\vspace{5mm}

In this section, we investigate the planar solution of a general non-degenerate theory $\Gamma$ in ${\cal C}_0$. 
As was shown in subsection \ref{matter}, the solution of a theory $\Gamma$ including {\bf f} can be obtained from the solution of another theory 
without {\bf f} through a minor modification. 
In the following, the diagram $\Gamma$ considered is assumed to have only {\bf bf}. 

\vspace{5mm}

Let us recall the technique used to solve a non-degenerate theory $\Gamma$ with two nodes. 
The resolvents $\{v^a(z)\}$ were obtained from the pull-back of functions $\{\Omega^a(u)\}$ defined on the $u$-plane. 
The relation between the coordinates $z$ and $u$ was given by the Abel-Jacobi map 
\begin{equation}
u(z)\ =\ \int_{p}^z\lambda. 
\end{equation}
This actually defines a map from the elliptic curve $E$ to its Jacobian $J(E)$ which is again a torus. 
The equations (\ref{2node_periodic}) suggest that $\Omega(z)$ would be given by a section of a vector bundle on $J(E)$. 
The vector bundle is specified by the monodromy matrices $I$ and $M_2M_1$ for the $A$-cycle and the $B$-cycle, respectively. 
By diagonalizing the monodromy matrices, the problem is reduced to the one for a pair of line bundles on $J(E)$. 
Since the theta functions are sections of some line bundles on $J(E)$ \cite{GH}, 
it is natural that $\Omega(z)$ can be written in terms of the theta functions. 

\vspace{5mm}

Now, consider a theory $\Gamma$ with many nodes. 
The equations (\ref{SP_homogeneous}), supplemented by the conditions for the absence of branch cuts, can be written as 
\begin{equation}
\omega(y_a^+,\xi)\ =\ \omega(y_a^-,\xi)M_a, \hspace{5mm} y_a\in[p_a,q_a], 
   \label{RH}
\end{equation}
where we have assumed $p_aq_a=1$, and 
\begin{equation}
M_a\ :=\ \left[ 
\begin{array}{ccccccc}
1 &           &    & n^{a1}    &    &           & \\
  & \ddots &    & \vdots   &    &           & \\
  &           & 1 & n^{a,a-1} &    &           & \\
  &           &    & -1         &    &           & \\
  &           &    & n^{a,a+1} & 1 &           & \\
  &           &    & \vdots   &    & \ddots & \\
  &           &    & n^{a,n_g} &    &           & 1
\end{array}
\right]. 
   \label{M_a}
\end{equation}
Note that each $M_a$ is diagonalizable and satisfy $M_a^2=I$. 

One may consider, instead of the elliptic curve $E$, a higher-genus Riemann surface $\Sigma$ defined by 
\begin{equation}
y^2\ =\ \prod_{a=1}^{n_g}(x-p_a)(x-q_a). 
\end{equation}
The genus of $\Sigma$ is $n_g-1$. 
The Abel-Jacobi map can be defined for higher-genus Riemann surfaces \cite{GH}. 
Let $\lambda^i$ $(i=1,\cdots,n_g-1)$ be a set of independent holomorphic 1-forms on $\Sigma$. 
Then, a set of integrals 
\begin{equation}
u^i(z)\ =\ \int_p^z\lambda^i, 
\end{equation}
where $p$ is a point on $\mathbb{C}$, defines a map from $\Sigma$ to its Jacobian $J(\Sigma)$ which is a torus of dimension $n_g-1$. 
This map is an embedding of $\Sigma$ into $J(\Sigma)$. 
The $A$-cycles and the $B$-cycles of $\Sigma$ are mapped to non-trivial cycles of $J(\Sigma)$. 

A possible generalization of the strategy used for the case $n_g=2$ would be to construct a vector bundle on $J(\Sigma)$ and then to take the 
pull-back of a section of the vector bundle. 
The vector bundle should be characterized by some monodromies around the non-trivial cycles on $J(\Sigma)$. 
In analogy with the case $n_g=2$, the monodromies for the cycles of $J(\Sigma)$ 
corresponding to the $A$-cycles of $\Sigma$ would be trivial, while those corresponding 
to the $B$-cycles would be given by matrices like $M_aM_{a-1}$. 
In this way, the equations (\ref{RH}) would define a homomorphism 
\begin{equation}
\rho\ :\ \pi_1(J(\Sigma))\ \to\ \mbox{GL}(n_g,\mathbb{C}), 
\end{equation}
that is, a representation of the fundamental group $\pi_1(J(\Sigma))$. 
The problem is that the matrices $\{M_aM_{a-1}\}$ do not commute among them. 
Since $\pi_1(J(\Sigma))$ is Abelian, such a representation $\rho$ cannot exist. 
Therefore, a generalization along this line is not successful. 

\vspace{5mm}

It would be still possible to regard $\omega(z,\xi)$ as a section of a vector bundle on $\Sigma$, not on $J(\Sigma)$, with monodromies obtained from 
$M_a$. 
The fundamental group $\pi_1(\Sigma)$ is a non-Abelian group which is generated by $2(n_g-1)$ letters 
$\alpha_1,\beta_1,\cdots,\alpha_{n_g-1},\beta_{n_g-1}$ subject to a single constraint 
\begin{equation}
\prod_{i=1}^{n_g-1}\alpha_i\beta_i\alpha_i^{-1}\beta_i^{-1}\ =\ 1. 
   \label{fundamental group}
\end{equation}
Note that from this description it is easy to notice that 
the fundamental group of a torus is Abelian. 
As mentioned above, there must exist a homomorphism from $\pi_1(\Sigma)$ to ${\rm GL}(n_g,\mathbb{C})$. 
A map defined by assigning the unit matrix to each $\alpha_i$ and a suitable product of $M_a$ to each $\beta_i$ apparently preserves the relations 
(\ref{fundamental group}), and therefore defines a homomorphism. 

The desired vector bundle on $\Sigma$ can be constructed as follows \cite{NS}. 
Let $\widetilde{\Sigma}$ be the universal cover of $\Sigma$. 
$\widetilde{\Sigma}$ can be regarded as a principal bundle $P$ over $\Sigma$ whose structure group is $\pi_1(\Sigma)$ \cite{Steenrod}. 
If an $n_g$-dimensional representation $\rho$ of $\pi_1(\Sigma)$ on $V=\mathbb{C}^{n_g}$ is given, then one can construct the associated vector bundle 
$P\times_\rho V$ which has the desired monodromy. 
The construction of the vector bundle is rather simple, however, the description of sections of the vector bundle does not seem to be tractable.

\vspace{5mm}

\subsection{Riemann-Hilbert problem}

\vspace{5mm}

It turns out that there is a more convenient reformulation. 
The equations (\ref{RH}) can be regarded as an implication of the existence of monodromies for $\omega(z,\xi)$ around the points $\{p_a,q_a\}$. 
The monodromy matrices at both $z=p_a,q_a$ are $M_a$ since $M_a^{-1}=M_a$ holds. 
Now the problem of solving the equations (\ref{RH}) becomes constructing a vector bundle on a punctured sphere 
\begin{equation}
{\cal D}\ := \ \mathbb{P}^1\backslash\{p_1,q_1,\cdots,p_{n_g},q_{n_g}\}, 
\end{equation}
whose (multi-valued) section $\omega(z)$ has the monodromies around the punctures prescribed above. 

As in the case of the higher-genus Riemann surface discussed above, the construction of the vector bundle is rather straightforward. 
The advantage of this reformulation is the relation to a well-studied mathematical problem, known as the Riemann-Hilbert problem, in its original form. 
This relation then provides us with a rather detailed description of sections of the vector bundle. 
A review of the Riemann-Hilbert problem can be found, for example, in \cite{Bolibrukh} and \cite{Bolibruch}, 
the latter of which also includes recent developments. 
A pedagogical introduction to the basics on analytic differential equations can be found in \cite{IY}. 

\vspace{5mm}

The Riemann-Hilbert problem in the original form discusses a relation between differential equations and monodromies exhibited by the solutions of them. 
Let ${\cal S}^F_n$ be the set of $n$-dimensional Fuchsian systems, a system of $n$ first-order differential equations with Fuchsian singularities only. 
Let $p_i$ $(i=1,\cdots,m)$ be points on $\mathbb{C}$. 
Then, a Fuchsian system in ${\cal S}^F_n$ whose Fuchsian singularities are at $\{p_i\}$ is of the form 
\begin{equation}
\frac{dy}{dz}\ =\ \sum_{i=1}^{m} \frac{A_i}{z-p_i} y
   \label{Fuchsian}
\end{equation}
where $y(z)$ is an $n$-vector-valued function and $\{A_i\}$ are constant $n\times n$ matrices. 
The Fuchsian system is non-singular at infinity if and only if 
\begin{equation}
\sum_{i=1}^{m}A_i\ =\ 0
\end{equation}
is satisfied. 
It is always assumed below that the point at infinity is a non-singular point. 

There exist $n$ independent solutions $\{y^l(z)\}$ $(l=1,\cdots,n)$ of (\ref{Fuchsian}) which are holomorphic on ${\cal D}$. 
As is familiar for the hypergeometric function, these solutions exhibit a monodromy when they are analytically continued around a singularity. 
It is convenient to consider the fundamental matrix 
\begin{equation}
Y(z)\ :=\ (y^1(z),\cdots,y^{n}(z)) 
\end{equation}
where the solutions $\{y^l(z)\}$ are regarded as column vectors. 
Note that $Y(z)$ is non-degenerate over ${\cal D}$. 
When $Y(z)$ is continued counterclockwise around a singularity $z=p_i$, it transforms as 
\begin{equation}
Y(z)\ \to\ Y(z)M_i 
\end{equation}
with a non-degenerate constant matrix $M_i$. 
Typically, $\log M_i$ is proportional to $A_i$ up to a similarity transformation. 

Let $\gamma_i$ denote a closed path, starting and ending at a chosen point, and encircling only one singularity $z=p_i$ once and counterclockwise. 
Suppose that the singularities are numbered such that the composition of the paths 
\begin{equation}
\gamma_1\circ\gamma_2\circ\cdots\circ\gamma_m
\end{equation}
is homotopically trivial on ${\cal D}$. 
Then, the monodromy matrices must satisfy 
\begin{equation}
M_m\cdots M_2M_1\ =\ I. 
   \label{trivial loop}
\end{equation}

This construction defines a map 
\begin{equation}
\phi_{\rm RH}\ :\ {\cal S}^F_n\ \to\ {\cal S}^M_n, 
\end{equation}
where ${\cal S}^M_n$ consists of sets of pairs $(p_i,M_i)$ satisfying the condition (\ref{trivial loop}). 
An element $\{(p_i,M_i)\}$ of ${\cal S}^M_n$ is called a monodromy data. 
When a monodromy data $\{(p_i,M_i)\}$ has a non-empty inverse image $\phi_{\rm RH}^{-1}(\{(p_i,M_i)\})$, 
it is said that the Riemann-Hilbert problem for this monodromy data has the positive solution. 

The Riemann-Hilbert problem was originally concerned with the surjectivity of $\phi_{\rm RH}$. 
At first, it was believed that $\phi_{\rm RH}$ is always surjective. 
However, various counterexamples have been found \cite{Bolibrukh}. 
Up to now, various sufficient conditions for the positive solvability of the Riemann-Hilbert problem have been found. 
One simple sufficient condition is the following \cite{Plemelj}: {\it if at least one of the monodromy matrix 
in a given monodromy data is diagonalizable, then the Riemann-Hilbert problem for this case 
has the positive solution. }
Another sufficient condition for the positive solvability of the Riemann-Hilbert problem is the irreducibility of the monodromy data 
\cite{Bolibruch'}\cite{Kostov}. 

These theorems can be proved by relating the problem of the existence of a Fuchsian system to the problem of the existence of a 
holomorphic vector bundle on 
$\mathbb{P}^1$ with a meromorphic connection. 
It can be shown \cite{IY} that, for any monodromy data, one can construct a holomorphic vector bundle on $\mathbb{P}^1$ with a meromorphic connection 
$\nabla$ such that 
a horizontal section $s$, a section satisfying $\nabla s=0$, exhibits the prescribed monodromy. 
The meromorphic connection can be also chosen such that it has only simple poles. 
If the holomorphic vector bundle is trivial, the equation $\nabla s=0$ in terms of a coordinate is equivalent to a Fuchsian system (\ref{Fuchsian}). 

\vspace{5mm}

When the Riemann-Hilbert problem has the positive solution for a given monodromy data, one may be able to obtain a matrix-valued function 
exhibiting the prescribed monodromy. 
This function is nothing but the fundamental matrix $Y_0(z)$ of the corresponding Fuchsian system. 

The structure of $Y_0(z)$ can be described explicitly \cite{Bolibrukh}. 
Let $E_i$ be a matrix related to the monodromy matrix $M_i$ as 
\begin{equation}
M_i\ =\ S_i^{-1}\exp\left( 2\pi iE_i \right)S_i 
   \label{log of monodromy}
\end{equation}
with a suitable non-degenerate matrix $S_i$. 
The matrix $E_i$ is chosen such that it is of the Jordan normal form, and its eigenvalues $\{\beta_i^l\}$ satisfy 
\begin{equation}
0\ \le \ \mbox{Re}(\beta_i^l)\ <\ 1. 
\end{equation}
Define a matrix-valued function $z^M$ for a matrix $M$ as 
\begin{equation}
z^M\ :=\ \exp\left( M\log z \right). 
\end{equation}
In terms of the matrices $E_i,S_i$, the local form of $Y_0(z)$ around $z=p_i$ can be written as 
\begin{equation}
Y_0(z)\ =\ U(z)(z-p_i)^{\Lambda_i}(z-p_i)^{E_i}S_i
   \label{local soln}
\end{equation}
where $U(z)$ is a matrix-valued function which is holomorphic and non-degenerate 
at $z=p_i$, and $\Lambda_i$ is a diagonal matrix whose components are integers. 
The matrix $\Lambda_i$ has to be chosen appropriately at each singularity such that $Y_0(z)$ is a section of a trivial holomorphic vector bundle. 

It can be shown \cite{Bolibruch2} that there exists a matrix-valued function $\Gamma(z)$ which is meromorphic on $\mathbb{P}^1$ and holomorphic 
and non-degenerate on 
${\cal D}\backslash\{\infty\}$ such that the product 
\begin{equation}
Y(z)\ :=\ \Gamma(z)Y_0(z)
\end{equation}
has the local form (\ref{local soln}) with $\Lambda_i$ absent at all singularities. 
Moreover, the function $Y(z)$ has the following simple form near infinity 
\begin{equation}
Y(z)\ =\ \mbox{diag}(z^{c_1},\cdots,z^{c_{n}})V(z) 
\end{equation}
where $\{c_l\}$ are integers and $V(z)$ is a matrix-valued function which is holomorphic and non-degenerate at infinity. 

The integers $\{c_l\}$ have the following geometric meaning \cite{Bolibruch}. 
Let $\widetilde{\cal D}$ be the universal cover of ${\cal D}$. 
As mentioned before, this can be regarded as a principal bundle on ${\cal D}$ whose structure group is the fundamental group $\pi_1({\cal D})$. 
One may construct a holomorphic ${\rm GL}(n,\mathbb{C})$ bundle $P$ associated to the principal bundle $\widetilde{\cal D}$. 
The representation $\rho:\pi_1({\cal D})\to {\rm GL}(n,\mathbb{C})$ necessary to construct $P$ is defined in terms of the monodromy matrices $\{M_i\}$. 
A section of this bundle defines a holomorphic connection $\nabla$ on $P$. 
It is possible to extend $P$ to a holomorphic ${\rm GL}(n,\mathbb{C})$ bundle on $\mathbb{P}^1$ with a meromorphic connection extending $\nabla$. 
Each extension is specified by a set of matrices $\{S_i,\Lambda_i\}$. 
The non-degenerate matrices $\{S_i\}$ are chosen such that the similarity transformation by $S_i$ makes $M_i$ upper-triangular. 
The other matrices $\{\Lambda_i\}$ are diagonal matrices with integer components. 
The canonical extension $P^\circ$ is defined as an extension of $P$ for which all $\Lambda_i$ are zero. 
It is known that $P^\circ$ does not depend on $\{S_i\}$. 
Let $F^\circ$ be the holomorphic vector bundle associated to $P^\circ$. 
The Birkhoff-Grothendieck theorem implies that 
any holomorphic vector bundle on $\mathbb{P}^1$ is holomorphically equivalent to a direct sum of line bundles. 
For the holomorphic vector bundle $F^\circ$, one finds 
\begin{equation}
F^\circ \ \cong\ {\cal O}(-c_1)\oplus\cdots\oplus{\cal O}(-c_{n}). 
\end{equation}
The integers $\{c_l\}$ are known as the splitting type of $F^\circ$. 

The sum of $c_l$ is determined by $E_i$ as \cite{Bolibruch3} 
\begin{equation}
\sum_{l=1}^{n}c_l\ =\ \sum_{i=1}^{m}\mbox{tr}(E_i). 
   \label{Chern number}
\end{equation}
Therefore, this sum can be determined solely from the monodromy data. 
An algorithm of calculating each $c_i$ is available \cite{Bolibruch2} once one knows the Fuchsian system explicitly from which one started. 

\vspace{5mm}

\subsection{The solution}

\vspace{5mm}

Let us return to the matrix model problem. 
The equations (\ref{RH}) imply that the monodromy data relevant to our case is ${\cal M}:=\{(p_a,M_a),(q_a,M_a)\}$. 
This monodromy data is consistent since 
\begin{equation}
\prod_{i=1}^{n_g}M(p_a)M(q_a)\ =\ I
\end{equation}
is trivially satisfied. 
As mentioned below (\ref{M_a}), any $M_a$ is diagonalizable. 
Therefore, the corresponding Riemann-Hilbert problem has the  positive solution, implying the 
existence of a matrix-valued function $Y(z)$ with the properties 
described in the previous subsection. 
It would be interesting to notice that the condition (\ref{exception}) is equivalent to the one for the reducibility of the monodromy. 
See Appendix \ref{irreducibility}. 

Each matrix $E_a$ can be chosen to be the following diagonal matrix 
\begin{equation}
E_a\ =\ \mbox{diag}\left( \frac12, 0, \cdots, 0 \right). 
   \label{E_a}
\end{equation}
This implies that the function $(z-p_a)^{E_a}$ is also diagonal, 
\begin{equation}
(z-p_a)^{E_a}\ =\ \mbox{diag}\left( \sqrt{z-p_a}, 1, \cdots, 1 \right). 
\end{equation}
The formula (\ref{Chern number}) implies 
\begin{equation}
\sum_{l=1}^{n_g}c_l\ =\ n_g. 
\end{equation}

\vspace{5mm}

To determine the values of $\{c_l\}$, 
let us first discuss the uniqueness of the solution of (\ref{SP_bipartite}). 
Suppose that there are two solutions $v(z)$ and $v'(z)$ written in the vector notation. 
The difference $\delta v(z):=v(z)-v'(z)$ satisfies the homogeneous equations (\ref{RH}). 
Therefore, the function $\delta v(z)$ has monodromies described by the monodromy data ${\cal M}$. 
Then a product 
\begin{equation}
r(z)\ :=\ \delta v(z)Y(z)^{-1} 
\end{equation}
has no monodromies, implying that the components of $r(z)$ are meromorphic functions on $\mathbb{P}^1$. 
Since both $v(z)$ and $v'(z)$ are assumed to be finite at the points $\{p_a,q_a\}$, so is $\delta v(z)$. 
The determinant of $Y(z)$ behaves like $\sqrt{z-p_a}$ or $\sqrt{z-q_a}$ near the singularities, and therefore, $Y(z)^{-1}$ cannot become singular enough 
to create poles in $r(z)$. 
It is concluded that $r(z)$ must be holomorphic on $\mathbb{C}$. 

The behavior of $r(z)$ near infinity is 
\begin{equation}
r(z)\ =\ ( \delta \tilde{v}^1(z)z^{-c_1},\cdots,\delta \tilde{v}^{n_g}(z)z^{-c_{n_g}} ), \hspace{5mm} \delta\tilde{v}(z)\ :=\ \delta v(z)V(z)^{-1}. 
\end{equation}
If all $\{c_l\}$ are positive, then this implies that $r(z)$ vanishes at infinity. 
Since $r(z)$ turns out to be holomorphic on $\mathbb{P}^1$ and vanishes at infinity, $r(z)$ must vanish, implying the uniqueness of the solution of 
(\ref{SP_bipartite}). 
On the other hand, if some of $\{c_l\}$ are zero or negative, then 
there are ambiguities of the solution which cannot be fixed by the conditions given so far. 

We found in section \ref{2node} that the solution is unique when $n_g=2$. 
This implies that in this case $c_1,c_2>0$. 
Since $c_1+c_2=2$ must be satisfied, one finds $c_1=c_2=1$. 
For all $n_g>2$, there exists a theory in ${\cal C}_0$ for which the existence and the uniqueness of the solution can be shown 
\cite{SS}. 
Therefore, for those theories, one must have 
\begin{equation}
c_1\ =\ \cdots\ =\ c_{n_g}\ =\ 1. 
   \label{splitting type}
\end{equation}
Since $\{c_l\}$ are integers, they would not change under continuous deformations of the holomorphic vector bundle $F^\circ$. 
If $\{n^{ab}\}$ are regarded as complex numbers, then one may be able to 
consider continuous deformations of $\{M_a\}$ which induce corresponding deformations 
of $F^\circ$. 
Note that the Riemann-Hilbert problem is still valid for such complex values of $\{n^{ab}\}$. 
Since any possible $\{M_a\}$ are connected to the monodromy matrices 
for a theory discussed in \cite{SS}, the splitting type $\{c_l\}$ would be (\ref{splitting type}) in general. 
This implies the uniqueness of the solution. 

Note that this conclusion would be invalid if there would exist 
a domain wall in the complexified parameter space $\{n^{ab}\}$ across which the values of $\{c_l\}$ 
jump. 
It would be interesting to clarify this issue. 

\vspace{5mm}

If the values of $\{c_l\}$ are given as (\ref{splitting type}), then it turns out that $Y(z)$ has a nice inversion property. 
To show this, consider $Y(z^{-1})$. 
This has the same monodromy property as $Y(z)$, due to the relation $p_aq_a=1$. 
This implies that a product $Y(z^{-1})Y(z)^{-1}$ has no monodromy. 
This product has a simple pole at the origin and a simple zero at infinity. 
This implies 
\begin{equation}
Y(z^{-1})Y(z)^{-1}\ =\ z^{-1}C 
\end{equation}
with a constant matrix $C$. 
Since the equality should hold at $z=1$, it is concluded that 
\begin{equation}
Y(z^{-1})\ =\ z^{-1}Y(z) 
\end{equation}
holds. 
At first sight, this would not be valid at $z=-1$. 
This is actually not that case since $Y(z)$ has branch cuts. 
As one takes $z=e^{i\theta}$ and changes the value of $\theta$ from 0 to $\pi$, one finds that $z$ 
and $z^{-1}$ pass through those cuts, resulting in the extra 
minus sign. 

The function $Y(z)$ does not change the monodromy if it is multiplied by a constant matrix from the left. 
Using this degree of freedom, we normalize $Y(z)$ such that $Y(0)$ is the unit matrix. 

\vspace{5mm}

Now, let us solve the equations (\ref{RH}). 
Suppose that there is a solution $\omega(z,\xi)$. 
Then, the product of this solution with $Y(z)^{-1}$ multiplied from the right has no monodromy. 
Therefore, they must be a vector-valued function whose components are meromorphic functions on $\mathbb{P}^1$. 
In fact, they are rational functions. 
Let $r(z,\xi)$ denote this vector-valued function. 
Then $\omega(z,\xi)$ can be written as 
\begin{equation}
\omega(z,\xi)\ =\ r(z,\xi)Y(z). 
\end{equation}

The function $r(z,\xi)$ is determined by requiring that $\omega(z,\xi)$ has the appropriate pole structure, as in section \ref{2node}. 
Recall that the function $Y(z)$ is defined such that $Y(z)$ is finite at the singularities $z=p_a,q_a$. 
Since we have assumed that $\omega(z,\xi)$ is finite at $z=p_a,q_a$, it turns out that $r(z,\xi)$ has poles only at $z=\pm\xi$. 
In addition, $r(z,\xi)$ should behave as $z^{-1}$ near infinity since $Y(z)$ diverges as $z$ while $\omega(z,\xi)$ was assumed to be finite. 
Therefore, $r(z,\xi)$ must have the following form 
\begin{equation}
r(z,\xi)\ =\ \frac1{z^2-\xi^2}\left[ a(\xi)z+b(\xi) \right]
\end{equation}
where $a(\xi),b(\xi)$ are $z$-independent row vectors. 
Let the required residues of $\omega(z,\xi)$ at $z=\pm\xi$ be denoted as 
\begin{equation}
\rho_{\pm}\ :=\ \mbox{Res}_{\pm\xi}\omega(z,\xi), 
\end{equation}
The vectors $a(\xi),b(\xi),c(\xi)$ are determined by the equations 
\begin{eqnarray}
a(\xi)\xi+b(\xi) &=& 2\rho_+\xi Y(\xi)^{-1}, 
   \label{RH_coefficients1} \\
a(\xi)\xi-b(\xi) &=& 2\rho_-\xi Y(-\xi)^{-1}. 
   \label{RH_coefficients2}
\end{eqnarray}
These equations can be solved easily, and the solution is 
\begin{eqnarray}
a(\xi) &=& \rho_+Y(\xi)^{-1}+\rho_-Y(-\xi)^{-1}, \\
b(\xi) &=& \xi\left[ \rho_+Y(\xi)^{-1}-\rho_-Y(-\xi)^{-1} \right]. 
\end{eqnarray}

Therefore, the solution can be written as 
\begin{equation}
\omega(z,\xi)\ =\ \left[ \frac1{z-\xi}\rho_+Y(\xi)^{-1}+\frac1{z+\xi}\rho_-Y(-\xi)^{-1} \right]Y(z). 
   \label{solution}
\end{equation}
As was argued above, this is the unique solution of the equations (\ref{RH}).

\vspace{5mm}

\subsection{Local theory of Fuchsian systems and 't~Hooft couplings}

\vspace{5mm}

It will be quite difficult, unless impossible, to write down the explicit form of $\omega(z,\xi)$ in terms of well-known functions. 
However, the solution (\ref{solution}) obtained above is not useless. 
It was found in subsection \ref{coupling} that, as long as we are interested in the divergent behavior of the 't~Hooft couplings, the only information 
necessary to extract it is the local behavior of the solution around the branch points. 
Fortunately, the local structure of $Y(z)$, which is the only non-trivial function in the solution (\ref{solution}), can be determined as (\ref{local soln}) 
with $\Lambda_a=0$ based on the local theory of Fuchsian systems \cite{Bolibrukh}. 

\vspace{5mm}

The 't~Hooft coupling $t=-v(0)$ is given as 
\begin{equation}
t\ =\ -\int_Cd\xi\left[ r(0,\xi)+\omega(0,\xi) \right], 
   \label{RH-t}
\end{equation}
where 
\begin{equation}
\omega(0,\xi)\ =\ \frac1{\xi}\left[ -\rho_+Y(\xi)^{-1}+\rho_-Y(-\xi)^{-1} \right]. 
\end{equation}
For the case $n_g=2$, the divergent behavior of interest appeared when one of the branch point 
approaches the origin, one of the endpoint of the 
integration contour $C$. 

Let us investigate the behavior of $t$ when a branch point, say $p_a$, approaches the origin. 
The determinant of $Y(\xi)$ near the point $\xi=p_a$ is 
\begin{equation}
\mbox{det}\,Y(\xi)\ =\ \mbox{det}\,U(\xi)\sqrt{\xi-p_a} 
\end{equation}
since in our case $\mbox{tr}(E_a)=\frac12$. 
Therefore, if $p_a$ is very close to the origin, then the behavior of $Y(\xi)^{-1}$ for small $\xi$ should be 
\begin{equation}
Y(\xi)^{-1}\ \sim\ \sqrt{\frac{-p_a}{\xi-p_a}}I, 
\end{equation}
due to the normalization $Y(0)=I$. 
This implies that the divergent behavior for a non-degenerate theory is quite similar to the one observed for the theory with $n_g=2$. 
In fact, the 't~Hooft coupling $t$ diverges as 
\begin{equation}
t\ \sim\ -(\rho_+-\rho_-)\log p_a
\end{equation}
as $p_a$ approaches the origin. 

In general, it would be possible that several branch points approach to the origin simultaneously. 
Even in such cases, the divergent behavior is similar. 
One may find that the integral (\ref{RH-t}) can be estimated roughly as 
\begin{equation}
t\ \sim\ -\int_\epsilon \frac{d\xi}{\xi}( -\rho_++\rho_- ) 
\end{equation}
where $\epsilon$ is determined by the parameters $\{p_a\}$. 
Therefore, in any case the divergence is logarithmic, and in particular, only the real part of $t$ may diverge.

\vspace{5mm}

\subsection{Isomonodromic deformations}

\vspace{5mm}

It has been shown that the solution for a non-degenerate 
theory with $n_g>2$ can be obtained, and the solution exhibits a similar divergent behavior of the 't~Hooft 
coupling $t$ as observed in section \ref{2node} for the theories with $n_g=2$. 
There was another observation for the theories with $n_g=2$. 
That is, the solution may not exist for a particular choice of the parameters. 
This might seem to be contradicting with the conclusion obtained in this section. 
Indeed, we found the solution of the equations (\ref{RH}) for any $n^{ab}$ and for any positions of the branch 
points. 
The existence of the solution was based on the positive solvability of the corresponding Riemann-Hilbert problem which was rigorously proved 
\cite{Plemelj}. 

The resolution comes from noticing that the solution in section \ref{2node} depends {\it analytically} on the positions of the branch points. 
The theorem in \cite{Plemelj} claims that, for any configuration of the singular points, there exists a Fuchsian system 
reproducing the prescribed monodromy. 
However, this does not claim that the family of the Fuchsian systems is {\it analytically} parametrized by the positions of the singularities. 
In fact, it is known \cite{JM1}\cite{JM2}\cite{Malgrange} 
that, although some of the Fuchsian systems form a continuous family parametrized by the positions, 
the parameter space is not 
the entire $\mathbb{C}^{m}$, where $m$ is the number of singularities of the Fuchsian system. 
For a particular choice the positions there is no Fuchsian system in the continuous family which reproduces the prescribed monodromy. 
The divergence of the 't~Hooft coupling $t$ should be related to this phenomenon. 

\vspace{5mm}

The relevant notion here is the isomonodromic deformations \cite{Schlesinger} of Fuchsian systems. 
One considers the following family of Fuchsian systems 
\begin{equation}
\frac{dy}{dz}\ =\ \sum_{i}\frac{A_i(a)}{z-a_i}y, 
   \label{RH_Fuchsian}
\end{equation}
where the matrices $\{A_i(a)\}$ depend on the positions $\{a_i\}$ of the singularities. 
The dependence of $A_i(a)$ on $\{a_i\}$ is chosen such that the monodromy realized by the solution of (\ref{RH_Fuchsian}) does not depend on $\{a_i\}$. 
A suitable change of $A_i(a)$ according to the continuous change of $a_i$ preserving the monodromy is called an isomonodromic deformation. 
The family of the Fuchsian systems (\ref{RH_Fuchsian}) should be the appropriate one 
for discussing the solution of the equations (\ref{RH}) since we would like to discuss physical quantities, for example the 't~Hooft coupling $t$, 
as a function of the positions of the branch points. 

The matrices $\{A_i(a)\}$ for an isomonodromic deformation are known to satisfy the Schlesinger equations 
\begin{equation}
dA_i(a)\ =\ -\sum_{j\ne i}\frac{[ A_i(a), A_j(a) ]}{a_i-a_j}d(a_i-a_j). 
   \label{Schlesinger}
\end{equation}
It was shown \cite{JM1}\cite{JM2}\cite{Malgrange} that the solution of this equation has movable singularities. 
A movable singularity is a singularity whose position depends on the initial condition for the solution, and 
therefore, the existence of such singularities is not easily anticipated from the equation. 
If a given set $\{a_i\}$ of values corresponds to the position of a movable singularity, 
then this implies that the Fuchsian system exhibiting a prescribed monodromy at the singularities $\{a_i\}$ 
does not exist within the continuous family of Fuchsian systems. 

For the case $n_g=2$, the above discussion on movable singularities can be made more explicit. 
It is known (see e.g. \cite{Iwasaki et al.}) that in this case the Schlesinger equations are reduced to the Painlev\'e VI equation 
\begin{eqnarray}
\frac{d^2y}{dx^2} 
&=& \frac12\left[ \frac1y+\frac1{y-1}+\frac1{y-x} \right]\left( \frac{dy}{dx} \right)^2-\left[ \frac1x+\frac1{x-1}+\frac1{y-x} \right]\frac{dy}{dx} \nonumber \\
& & +\frac{y(y-1)(y-x)}{x^2(x-1)^2}\left[ \alpha+\beta\frac x{y^2}+\gamma\frac{x-1}{(y-1)^2}+\delta\frac{x(x-1)}{(y-x)^2} \right]. 
   \label{P_VI}
\end{eqnarray}
Note that the relation between matrices in the Fuchsian system and the parameters $\alpha,\beta,\gamma,\delta$ in (\ref {P_VI}) 
was given, for example in \cite{PVI}, for the case when the monodromy matrices belong to $\mbox{SL}(2,\mathbb{C})$, but in our case this is not since 
$\mbox{det}\,M_a=-1$. 
Actually, this is not a big problem since any monodromy in $\mbox{GL}(n,\mathbb{C})$ can be decomposed into a one-dimensional monodromy and a 
monodromy in $\mbox{SL}(n,\mathbb{C})$, and the one-dimensional monodromy problem can be easily solved. 
In the case when all monodromy matrices belong to $\mbox{SL}(2,\mathbb{C})$, the parameters $\alpha,\beta,\gamma,\delta$ are determined 
by the eigenvalues $\pm\frac12\alpha_i$ of the matrices $\{A_i\}$ in the Fuchsian system as 
\begin{equation}
\alpha\ =\ \frac{(\alpha_1-1)^2}2, \hspace{5mm} \beta\ =\ -\frac{\alpha_2^2}2, \hspace{5mm} \gamma\ =\ \frac{\alpha_3^2}2, \hspace{5mm} 
 \delta\ =\ \frac12-\frac{\alpha_4^2}2. 
\end{equation}
The matrix $A_i$ is equal to a matrix $E_i$, which is related to the monodromy matrix as in (\ref{log of monodromy}), up to a similarity transformation. 
In our case, one finds $\alpha_i=\frac14$, implying 
\begin{equation}
\alpha\ =\ \frac18, \hspace{5mm} \beta\ =\ -\frac18, \hspace{5mm} \gamma\ =\ \frac18, \hspace{5mm} \delta\ =\ \frac38. 
\end{equation}

Quite remarkably, even though the Painlev\'e VI equation (\ref{P_VI}) is highly non-linear, a two-parameter family of solutions of this equation is known 
\cite{Hitchin} 
for this choice of parameters. 
It is written in terms of theta functions. 
The explicit form is given in the following parametric form 
\begin{eqnarray}
x(\tau) &=& \frac{\vartheta_3(0)^4}{\vartheta_4(0)^4}, \\
y(\tau) 
&=& \frac{\vartheta_1'''(0)}{3\pi^2\vartheta_4(0)^4\vartheta_1'(0)}+\frac13\left[ 1+\frac{\vartheta_3(0)^4}{\vartheta_4(0)^4} \right] \nonumber \\
& & +\frac{\vartheta_1'''(\nu)\vartheta_1(\nu)-2\vartheta_1''(\nu)\vartheta_1'(\nu)+4\pi ic_1( \vartheta_1''(\nu)\vartheta_1(\nu)-\vartheta_1'(\nu)^2 )}
      {2\pi^2\vartheta_4(0)^4\vartheta_1(\nu)( \vartheta_1'(\nu)+2\pi ic_1\vartheta_1(\nu) )}, 
\end{eqnarray}
where the modulus $\tau$ of the theta function is the modulus of the elliptic curve defined by 
\begin{equation}
y^2\ =\ \prod_{i=1}^4(x-a_i), 
\end{equation}
and $\nu$ here is related to the parameters $c_1,c_2$ as 
\begin{equation}
\nu\ =\ c_1\tau+c_2. 
\end{equation}
It is apparent that $y(x)$ has a pole at $x$ for which 
\begin{equation}
c_1\tau+c_2\ \in\ \mathbb{Z}+\mathbb{Z}\tau. 
\end{equation}
This is quite similar to the condition (\ref{2node_no solution}) for which there is no solution of the equations (\ref{SP_homogeneous}), and for which 
the 't~Hooft couplings diverge. 
Exactly speaking, the modulus $\tau$ here corresponds to $\tau^{-1}$ in section \ref{2node}, but this does not make any essential modification.

\vspace{1cm}

\section{Discussion}  \label{discuss}

\vspace{5mm}

We have investigated the planar solution of various Chern-Simons-matter theories. 
Under the assumption of the symmetric distributions of the eigenvalues, one can reduce the problem of solving a generic Chern-Simons-matter 
theory to the one of solving the corresponding theory including only fundamental and bi-fundamental hypermultiplets. 
Due to this planar relations, it was shown that 
many of the theories in the family ${\cal C}$ defined in section \ref{saddle} may admit a systematic analysis of the planar resolvents. 
As a result, we obtained an integral representation of the resolvents whose integrands are written in terms of the solution of a Fuchsian system, 
a set of first-order differential equations, as long as the theory is non-degenerate, that is, 
the condition (\ref{exception}) is not satisfied. 
Local properties of the solution of Fuchsian systems have been studied so far. 
The results can be used to clarify the properties of physical observables 
of the Chern-Simons-matter theories. 
It was found that the properties of the non-degenerate theories are quite different from those of ABJM theory. 
For example, the vevs of Wilson loops do not grow exponentially as the (physical) 't~Hooft couplings grow. 
These theories are rather similar to the Chern-Simons-adjoint theories investigated in \cite{Suyama:2012uu}. 
For the case when the gauge group is the product of two ${\rm U}(N_a)$ factors, the integrands for the resolvents can be written more explicitly in terms 
of theta functions, and the properties of the solution can be analyzed in detail. 

An interesting result obtained in this paper is the condition (\ref{exception}). 
According to the results in section \ref{reform}, it would be expected that an interesting large 't~Hooft coupling limit exists only 
for degenerate theories. 
Note that the Chern-Simons-matter theories satisfying the ``no long-range force'' condition discussed in \cite{Gulotta:2012yd} 
also satisfy (\ref{exception}). 
From the point of view of the matrix model, (\ref{exception}) 
is a rather technical condition which shows the inability to apply a particular technique developed in this paper. 
It would be very interesting to clarify the meaning of this condition in the context of quantum field theory. 
Since this condition seems to be a necessary condition for Chern-Simons-matter theories to have simple gravity duals, the understanding of this 
condition would provide us with an insight on the mechanism of the emergence of a dual geometry in the context of AdS/CFT correspondence. 
The condition (\ref{exception}) should be also important to understand the large $N$ behavior of the free energy which scales as $N^{\frac32}$ for 
ABJM theory \cite{Drukker:2010nc} in terms of quantum field theory. 
Since such a behavior is not universal for general Chern-Simons-matter theories, the explanation should take into account some effects of a particular 
choice of the matter contents. 

It would be also interesting to classify the solutions of (\ref{exception}) and to investigate the properties of the corresponding theories. 
It was shown in subsection \ref{homogeneous} that the solutions are not exhausted by $n^{ab}$ corresponding to ADE affine Dynkin diagrams. 
To analyze those degenerate theories, 
one has to invent a new technique for solving the saddle-point equations for those theories. 
Probably, some of those theories can be solved more completely by using the Fermi gas approach \cite{Marino:2011eh}. 
It would be interesting to specify the subset of those degenerate theories to which the Fermi gas approach is applicable. 

\begin{figure}[tbp]
\begin{center}
\includegraphics{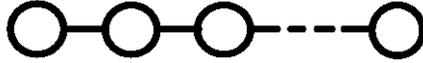}
\end{center}
\caption{
The Chern-Simons-matter theory corresponding to this diagram can be solved by the technique in this paper. 
This is realized by a linear array of D5-branes and NS5-branes with D3-branes suspended among them. 
}
\label{linear diagram}
\end{figure}

It is usually quite helpful to find a brane realization of a Chern-Simons-matter theory for understanding its properties. 
Of course, one should not expect that all the theories in the family ${\cal C}$ might have such a brane realization. 
Since many of the theories with known gravity duals satisfy (\ref{exception}), one might anticipate that brane realization might not be available for 
non-degenerate theories. 
However, it would be interesting to notice that there exist some non-degenerate Chern-Simons-matter theories with brane realizations. 
An example is the theory corresponding to a linear diagram depicted in Figure \ref{linear diagram}. 
With the help of the knowledge on the brane dynamics, one may be able to investigate such a theory in more detail \cite{SS}. 

An interesting physical issue for Chern-Simons-matter theories is supersymmetry breaking. 
For simple theories, it is possible to investigate the condition for when the supersymmetry is spontaneously broken 
\cite{Witten:1999ds}\cite{Bergman:1999na}\cite{Ohta:1999iv}\cite{Suyama:2012kr}. 
Recently, it is argued that the free energy (or the partition function) can be regarded as a kind of an order parameter of 
supersymmetry breaking \cite{Morita:2011cs}. 
For example, for pure Chern-Simons theory, the partition function \cite{Kapustin:2009kz}
\begin{equation}
Z\ \propto \prod_{m=1}^N\left( \sin\frac{\pi m}{k} \right)^{N-m}
\end{equation}
vanishes when the supersymmetry is broken, that is, $N>k$. 
It would be interesting to evaluate the free energy using the planar resolvent obtained in this paper, and systematically investigate singular behaviors 
of the free energy to check whether it is really related to supersymmetry breaking. 

The possible dual to a Chern-Simons-matter theory is not only a gravity theory, but it may be a higher-spin theory \cite{Chang:2012kt}. 
It would be interesting to find a criterion by which one can judge whether a given Chern-Simons-matter theory may be dual to a higher-spin theory.

\vspace{2cm}

{\bf \Large Acknowledgements}

\vspace{5mm}

I would like to thank Soo-Jong Rey and Igor Shenderovich for valuable discussions and comments. 
This work was supported in part by the National Research Foundation grants NRF-2005-0093843, NRF-2010-220-C0000, NRF-2012KA1A9055, 
NRF-20120007209.

\appendix

\vspace{1cm}

\section{The Abel-Jacobi map}  \label{Abel-Jacobi}

\vspace{5mm}

Let $\lambda_0$ be the following holomorphic 1-form 
\begin{equation}
\lambda_0\ :=\ \frac{dz}{\sqrt{(z^2-p^2)(z^2-q^2)}}. 
\end{equation}
on the elliptic curve $E$ defined by 
\begin{equation}
y^2\ =\ (x^2-p^2)(x^2-q^2). 
\end{equation}
The integrals of $\lambda_0$ 
\begin{equation}
\pi_A\ :=\ \int_A\lambda_0, \hspace{5mm} \pi_B\ :=\ \int_B\lambda_0 
\end{equation}
over the $A$-cycles and the $B$-cycles in Figure \ref{cycles} satisfy the second Riemann relation 
\begin{equation}
\mbox{Re}(\pi_A)\mbox{Im}(\pi_B)-\mbox{Re}(\pi_B)\mbox{Im}(\pi_A)\ >\ 0. 
\end{equation}
Usually, a normalized 1-form is used for which $\pi_A=1$ such that $\mbox{Im}(\pi_B)>0$. 
In section \ref{2node}, we instead employ another normalization 
\begin{equation}
\lambda\ :=\ \frac1{\pi_B}\lambda_0, 
\end{equation}
for which $\mbox{Im}(\pi_A)<0$. 

The map 
\begin{equation}
u(z)\ :=\ \int_p^z\lambda 
\end{equation}
defined in subsection \ref{general solution} maps several points in the $z$-plane as follows, 
\begin{equation}
(-q,-p,0,p,q,\infty)\ \mapsto\ \left( \frac12-\frac\tau2, \frac12, \frac14, 0, -\frac\tau2, \frac14-\frac\tau2 \right), 
\end{equation}
and a half of the $z$-plane is mapped onto a parallelogram formed by the above points. 
The other half is mapped to another parallelogram. 

It is obvious from the definition that 
\begin{equation}
u(\widetilde{z})\ =\ -u(z)
\end{equation}
holds, where $\widetilde{z}$ is a point on the second Riemann sheet. 
It is also easy to show that 
\begin{equation}
u(-z)\ =\ \int_{p}^{-z}\lambda\ =\ -u(z)+\frac12. 
\end{equation}
Combining these two relations, one obtains 
\begin{equation}
u(\widetilde{z})\ =\ u(-z)-\frac12. 
\end{equation}

The $u$-coordinate of $z^{-1}$ should be given with care. 
Naively, 
\begin{equation}
\int^{z^{-1}}_p\lambda\ =\ \int^z_p\lambda+\int_q^p\lambda 
\end{equation}
suggests 
\begin{equation}
u(z^{-1})\ =\ u(z)+\frac\tau2. 
\end{equation}
A precise definition would require the choice of a contour, and also the determination of which Riemann sheet $z^{-1}$ lies on. 
We instead define $z^{-1}$ such that the above relation holds. 
Note that this implies $1^{-1}=\widetilde{1}$ etc.

\vspace{1cm}

\section{Coefficients $c_i(\xi)$}  \label{coefficients}

\vspace{5mm}

The coefficients $\{c_i(\xi)\}$ in subsection \ref{general solution} are determined by the following linear equation 
\begin{equation}
\left[ 
\begin{array}{cccc}
1 & \xi & f_3(\xi) & f_4(\xi) \\
1 & -\xi & f_3(-\xi) & f_4(-\xi) \\
1 & \xi & f_4(-\xi) & f_3(-\xi) \\
1 & -\xi & f_4(\xi) & f_3(\xi) \\
\end{array}
\right]\left[ 
\begin{array}{c}
c_1(\xi) \\ c_2(\xi) \\ c_3(\xi) \\ c_4(\xi) 
\end{array}
\right]\ =\ \frac{2i\xi}{\sin\pi\nu}\left[ 
\begin{array}{c}
\kappa_2G_+ \\ -\kappa_1G_- \\ \kappa_2G_- \\ -\kappa_1G_+
\end{array}
\right], 
\end{equation}
where 
\begin{equation}
G_\pm\ :=\ e^{\pm\frac12\pi i\nu}G(\pm\xi)^{-1}. 
\end{equation}

The solution is 
\begin{eqnarray}
c_1(\xi) &=& \frac{i(\kappa_1-\kappa_2)}{4z_\nu\sin\pi\nu}\frac1{y(\xi)}(A_-G_+-A_+G_-), \\
c_2(\xi) &=& -\frac{i(\kappa_1+\kappa_2)}{4z_\nu\sin\pi\nu}\frac1{y(\xi)}(B_-G_+-B_+G_-), \\
c_3(\xi) &=& \frac{i}{4z_\nu\sin\pi\nu}\frac1{y(\xi)}(\kappa_2C_++\kappa_1C_-)(G_+-G_-), \\
c_4(\xi) &=& -\frac{i}{4z_\nu\sin\pi\nu}\frac1{y(\xi)}(\kappa_1C_++\kappa_2C_-)(G_+-G_-), 
\end{eqnarray}
where 
\begin{eqnarray}
A_\pm &=& 2z_\nu[ z_\nu y(z_\nu)\pm\xi y(\xi) ], \\
B_\pm &=& 2[ \xi y(z_\nu)\pm z_\nu y(\xi) ], \\
C_\pm &=& (\xi\pm z_\nu)(\xi^2-z_\nu^2). 
\end{eqnarray}

These coefficients are finite at $\xi=0$. 
Recalling that $G(z)$ has a simple pole at infinity, $c_i(\xi)$ are also finite at $\xi=\infty$. 
This implies that the integral (\ref{ansatz}) are well-defined.

\vspace{1cm}

\section{Irreducibility of the monodromy}  \label{irreducibility}

\vspace{5mm}

The monodromy representation discussed in section \ref{reform} is the one defined by the matrices $\{M_a\}$ given as (\ref{M_a}). 
This is irreducible if there is no proper subspace of $\mathbb{C}^{n_g}$ which is fixed by all $M_a$. 

Assume that there exists a proper subspace $V\subset \mathbb{C}^{n_g}$ such that 
\begin{equation}
V\cdot M_a\ \subset\ V. 
\end{equation}
Let $m<n_g$ be the dimension of $V$. 
One may choose a basis of $V$. 
The components of the basis vectors form an $m\times n_g$ matrix 
\begin{equation}
\left[ 
\begin{array}{cccc}
c^1_1, & c^2_1, & \cdots & c^{n_g}_1 \\
\vdots & \vdots & & \vdots \\
c^1_m, & c^2_m, & \cdots & c^{n_g}_m 
\end{array}
\right]. 
\end{equation}
The basis can be chosen such that all $m\times m$ minors are non-degenerate. 
In fact, one choice of such a basis corresponds to a point of the Grassmannian ${\rm Gr}_{m,n_g}(\mathbb{C})$ 
contained in the intersection of all the standard 
coordinate patches, which is obviously non-empty. 

Let $v$ be a vector in $V$. 
It is parametrized by $m$ variables $\{t^i\}$ as 
\begin{equation}
v\ =\ (c^1_it^i, \cdots, c^{n_g}_it^i). 
\end{equation}
Let us fix a value of $a$. 
By assumption, $vM_a$ is in $V$. 
Therefore, there is another set of parameters $\{\tilde{t}^i\}$ such that 
\begin{equation}
vM_a\ =\ (c^1_i\tilde{t}^i, \cdots, c^{n_g}_i\tilde{t}^i). 
\end{equation}
This and the explicit form (\ref{M_a}) of $M_a$ implies that, for $b\ne a$, 
\begin{equation}
c^b_i( t^i-\tilde{t}^i )\ =\ 0. 
\end{equation}
One can choose $m$ of these equations for which $c^b_i$ is non-degenerate. 
These equation implies $t^i=\tilde{t}^i$ for all $i$. 

The remaining equations are 
\begin{equation}
\sum_b n^{ab}c^b_it^i-2c^a_it^i\ =\ 0. 
\end{equation}
There exists a non-trivial solution for $\{t^i\}$ if and only if 
\begin{equation}
\mbox{det}\left[ 2\delta^{ab}-n^{ab} \right]\ =\ 0. 
\end{equation}
Therefore, for a non-degenerate theory for which this is not satisfied, the monodromy representation is irreducible.

\end{document}